\documentstyle[12pt,epsf]{article}
{}{}
{}{}
\begin{document}                

\begin{minipage}{12cm}
\hspace*{8.5cm} {\bf Preprint JINR}\\
\hspace*{9.cm}{\bf E2-2001-246}\\
\hspace*{9.0cm}{\bf Dubna, 2001}
\end{minipage}

\vspace*{2.cm}

\begin{center}
{\large \bf
Asymptotic properties of high-$p_T$ particle production\\[0.03cm]
in hadron-hadron, hadron-nucleus and nucleus-nucleus \\[0.15cm]
collisions  at high energies}
\vskip 1.cm {\bf M. V. Tokarev\footnote{E-mail: tokarev@sunhe.jinr.ru}}
\vskip 0.5cm
{\it
Laboratory of High Energies,
Joint Institute for Nuclear Research,\\
141980, Dubna, Moscow region, Russia
}

\vskip 1cm
\begin{abstract}
The concept of $z$-scaling reflecting the general features of internal particle substructure, constituent
interaction and mechanism of real particle formation is reviewed. Experimental data on the cross sections
obtained at ISR, SpS and Tevatron are  used in the analysis. The properties of data $z$-presentation, the
energy and angular independencies, the power law, $A$- and $F$-dependencies, are discussed. It is argued that
the properties reflect the fundamental symmetries such as self-similarity, locality, fractality and
scale-relativity. The use of $z$-scaling to search for new physics phenomena in hadron-hadron,  hadron-nucleus
and nucleus-nucleus collisions is suggested. The violation of $z$-scaling characterized by the change of the
fractal dimension is considered as  a new and complimentary signature of nuclear phase transition.
\end{abstract}

\vskip 1cm

{\it
 Presented at the 6th International Workshop\\
 "Relativistic Nuclear Physics: from Hundreds of MeV to TeV",\\
September 10-16, 2001, Varna, Bulgaria
}

\end{center}

\vskip 1.5cm
\newpage

{\section{Introduction}}

Asymptotic properties of particle formation are believed to reveal themselves more clearly at high energy
$\sqrt s $ and transverse momenta. It is considered also that partons produced in hard scattering retain
information about primary collision during hadronization. The mechanism of particle formation can be modified
by nuclear environment and can be sensitive to the phase transition as well. Therefore, the  features of
single inclusive particle spectra of hadron-hadron, hadron-nucleus and nucleus-nucleus collisions  are of
interest to search for new physics phenomena, for example the phase transition of nuclear matter, at extremal
conditions and a quantitative test of theory.

One of the methods to study the properties of particle formation in
nuclear matter is to search for the violation of
 known regularities , e.g. the Bjorken, Feynman scaling laws and quark counting rules \cite{matveev,brodsky},
established in elementary collisions ($l-p$, $p-p$, $\bar p-p$, etc.).

In the report the general concept of the new scaling, $z$-scaling, \cite{Z96,Z97,Z99,Z01} is reviewed. The
scaling reflects the properties of particle formation over a high $p_{T}$ range in $h-h$, $h-A$ and $A-A$
collisions at high energies.

The scaling  was proposed in \cite{Z96} to describe the feature
of charged hadron produced in $p-p$ and $\bar p-p$ collisions.
The idea of the scaling  was developed for the analysis of direct photon
production in $p-p$ \cite{ppg}, $\bar p-p$ \cite{barppg} and $p-A$ \cite{pag}
collisions. The scaling
properties of jet production in $\bar p-p$ and $p-p$ collisions
were analyzed in \cite{Dedovich}.
 The scaling features of $\pi^{0}$-meson production in
 $p-p$, $p-A$ and $A-A$  collisions were established in  \cite{Rog1,Rog2}
 and charged and neutral hadrons produced in $\pi-p$ and $\pi-A$ collisions were studied  in \cite{Tok01}.
  The $z$-scaling relevance to
  the fractal structure of space-time itself was discussed in \cite{Zb,Z00}.

  The general concept of the scaling is based on the fundamental principles
  of self-similarity, locality, fractality and scale-relativity.
  The first one reflects the dropping   of certain dimensional quantities or
  parameters out of the physical picture of the interactions. The second
  principle concludes that the momentum-energy
  conservation law is locally valid  for the interacting constituents.
  The fractality principle says that both
  the structure of interacting particles and their formation
  are self-similar revealing properties of fractals at any
  scale.
  The fourth one, the  scale relativity principle, states that the
  self-similar and fractal substructures in the interaction obey
  relativistic principle concerning various scales \cite{Nottale,Z99}.

  As shown in \cite{Z96,Z99,Z97,Z01},
  the $z$-presentation of experimental data can be obtained
  using the experimental quantities. They are the inclusive cross
  section $Ed^3\sigma/dq^3$ and the multiplicity density of charged particles
  $\rho(s,\eta)$.
  The scaling function  $\psi(z)$ is found
  to be independent of the center-of-mass energy ${\sqrt s}$ and the angle
  ${\theta}$ of the produced particle over a
  wide kinematic range. The function  $\psi(z)$ describes the
  probability density to form a particle with  a
  formation length $z$. The scaling variable $z = z_0  \epsilon^{-\delta}$
  reveals the property  of a fractal measure,
  where $\epsilon $ is the scale resolution, and
  $\delta$ was interpreted as the  fractal dimension
  in the particle formation process.
  It was shown \cite{Z01} that, in the
  framework of $z$-presentation,  the A-dependence
  of high $q_T$ hadron production  is described by the function $\alpha$
  depending on the single parameter, the atomic weight $A$.
  The existence of the scaling  and its properties
  is assumed to reflect the fundamental features of
  particle structure, constituent interaction and particle
  production such as self-similarity, locality,
  fractality and scale-relativity.

{\section{$Z$-scaling}}

The idea of $z$-scaling \cite{Z96} is based on the assumptions that inclusive particle distribution of the
process (\ref{eq:r1}) at high energies can be described in terms of the corresponding kinematic
characteristics
\begin{equation}
P_{1}+P_{2} \rightarrow q + X \label{eq:r1}
\end{equation}
of the exclusive sub-process \cite{Stavinsky} written in the symbolic form (\ref{eq:r2})
  \begin{equation}
  (x_{1}M_{1}) + (x_{2}M_{2}) \rightarrow m_{1} +
  (x_{1}M_{1}+x_{2}M_{2} + m_{2})
  \label{eq:r2}
  \end{equation}
and that the scaling function depending on a single variable $z$ exists and is expressed via the dynamic
quantities, invariant inclusive  cross section $Ed^3\sigma/{dq^{3}}$ of the process (\ref{eq:r1}) and particle
multiplicity density $\rho(s,\eta)$. The kinematic quantities of the process (\ref{eq:r1}) are $P_1,P_2,q$ and
$M_{1},  M_{2}, m_1$,
  the momenta and masses of the colliding objects (hadron, nuclei)
  and inclusive particles, respectively.
   The parameter $m_{2}$ is introduced  to satisfy the  internal conservation
  laws (for isospin, baryon number,
  and strangeness). The $x_{1}$ and $x_{2}$ are the scale-invariant
  fractions of the incoming four-momenta
  $P_{1}$ and $P_{2}$. They determine the
  minimum energy, which
  is necessary for the production of the secondary particle with
  the mass $m_1$ and the four-momentum $q$.

\vskip 0.5cm

  {\subsection{Fractions $x_1$ and $x_2$}}

  The elementary parton-parton collision is considered  as a binary
  sub-process which satisfies the condition

  \begin{equation}
  (x_{1}P_{1} + x_{2}P_{2} - q)^{2} = (x_{1}M_{1} + x_{2}M_{2} +
  m_{2})^{2}.
  \label{eq:r5}
  \end{equation}
  The equation reflects minimum recoil mass hypothesis in the
  elementary sub-process. To connect kinematic and structural
  characteristics of the interaction, the coefficient
  $\Omega$ is introduced. It is chosen in the form
  \begin{equation}
  \Omega(x_1,x_2) = m(1-x_{1})^{\delta_1}(1-x_{2})^{\delta_2},
  \label{eq:r8}
  \end{equation}
  where $m$ is a mass constant and $\delta_1$ and $\delta_2$
  are factors relating to the fractal structure of
  the colliding objects \cite{Z99}.
  The fractions $x_{1}$ and
  $x_{2}$  are determined  to maximize the value of $\Omega(x_1,x_2)$,
  simultaneously fulfilling condition (\ref{eq:r5})
  \begin{equation}
  {d\Omega(x_1,x_2)/ dx_1}|_{x_2=x_2(x_1)} = 0.
  \label{eq:r9}
  \end{equation}
  The variables
  $x_{1,2}$ are equal to unity along the phase space limit and
  cover the full phase space accessible at any
  energy.

\vskip 0.5cm

{\subsection{Scaling function $\psi(z)$ and variable $z$}}

The scaling function $\psi(z)$ is written in the form \cite{Z99}
 \begin{equation}
 \psi(z) = - \frac{\pi s_A}{\rho_A(s,\eta) \sigma_{inel}}J^{-1}
 E\frac{d\sigma}{dq^{3}}.
 \label{eq:r20}
 \end{equation}
Here $\sigma_{inel}$ is the inelastic cross section,
$s_A \simeq s \cdot A$ and $s$ are the center-of-mass energy
 squared of the corresponding $ h-A $
 and $ h-N $ systems, $A$ is the atomic weight and
$\rho_A(s,\eta)$ is the average particle multiplicity density. The factor $J$ is the known function of the
kinematic variables, the momenta and masses of the colliding and produced particles \cite{Z99}.

 We would like to emphasize that the function  $\psi(z)$
 depends on a single scaling variable $z$.
 The existence of such a solution is not evident in advance.

The expression (\ref{eq:r20}) relates the differential
 cross section for the production of  the inclusive particle $m_{1}$
 and the average particle  multiplicity density $\rho_A(s,\eta)$
 with the scaling function $\psi(z)$.
The function is normalized as
\begin{equation}
\int_{z_{min}}^{\infty} \psi(z) dz = 1.
\label{eq:b6}
\end{equation}
The equation allows us to give the physical meaning
of the function $\psi$ as a probability density to form
a particle  with the corresponding value of the variable $z$.

  In accordance with the ansatz suggested in \cite{Z99}
 the variable $z$ is taken in the form
 (\ref{eq:r28})
 as a simple physically meaningful variable reflecting
   self-similarity and fractality as a general
    pattern of hadron production at high energies
\begin{equation}
z = \frac{ \sqrt{ {\hat s}_{\bot} }} {\Omega \cdot \rho_A(s) }.
\label{eq:r28}
\end{equation}
Here $\sqrt{ {\hat s}_{\bot} }  $
is the minimal transverse energy of
 colliding constituents necessary to produce a real hadron in the reaction
(\ref{eq:r1}). The factor $\Omega$ is  given by (\ref{eq:r8})
 and   $\rho_A(s) =\rho_A(s, \eta)|_{\eta=0}$.
The transverse energy consists of two parts
which represent the transverse energy of the inclusive particle
and its recoil.
 The form of $z$ determines its variation range $(0,\infty) $.
 These values are scale independent and
 kinematically accessible at any energy.

  One of the features of the procedure to construct $\psi(z)$   and $z$
described above  is the joint use of the experimental quantities  characterizing hard ($Ed^3\sigma/{dq^{3}}$)
and soft ($\rho_A(s, \eta)$) processes of particle interactions.
  Therefore, there is a real problem
  for a theoretical description of $z$-scaling in the
  framework of perturbative QCD. We would like to note that
  $z$-construction is not direct mathematical consequence of
  parton model of strong interaction but it is a new
  self-similarity pattern motivated by parton-parton and string-like
  scenarios of particle interactions.

    Let us consider the definition of the variable
   $z=\sqrt{\hat s_{\bot}}/(\Omega\rho_A)$  more closely
   and clarify its physical meaning.
  The value $ \sqrt {\hat s_{\bot}}$ is the minimal transverse energy of
  colliding constituents necessary to produce a real hadron in the reaction
  (\ref{eq:r1}). It is assumed that two point-like and massless
  elementary constituents interact with each other in the initial state and
  convert into real hadrons in the final state. The conversion is not
  instant process and is called  hadronization or particle formation.
  The microscopic space-time picture of the hadronization is not
  understood enough at present time.
  We assume that number of hadrons produced
  in the hard interaction of constituents
  is proportional to $\rho_A$. Therefore the value
  $ \sqrt {\hat s_{\bot}}/\rho_A$ corresponds to the energy density
  per one hadron produced in the sub-process.
  The factor $\Omega $ is relative number of all initial
  configurations containing the constituents which carry the
  momentum fractions $x_1$ and $x_2$. This factor thus represents
  a tension in the considered sub-system with respect to the
  whole system.
  Taking into account the qualitative scenario of hadron formation
  as a conversion of a point-like constituent into a real hadron
  we interpreted the variable $z$ as particle formation length.

\vskip 0.5cm

{\subsection{Fractality}}

  Fractality in particle and nuclear physics concerns the internal
  structure of particles, their interactions and formation of real particle.
  It is manifested  by their self-similarity on any scale.
  This general principle reflects the existence of power law dependencies of
  the corresponding quantities  \cite{Nottale,Z99}.
  In our case
 the quantity  $\Omega$, given by (\ref{eq:r8}),
 connects the kinematic and fractal characteristics of
 the interaction and is described by the power law. As it will be shown below the scaling function
 $\psi(z)$
reveals  the power behavior in the asymptotic region too.
 The factors $\delta_1$ and $\delta_2$ are
  fractal dimensions of the colliding objects.
 The fractal structure itself is defined by the structure
 of the interacting constituents, which is not an elementary one either.
 In this scheme, high energy hadron-hadron, hadron-nucleus and
 nucleus-nucleus interactions are considered as interactions of fractals.

In the case of collisions of asymmetric objects,
the approximation for the measure $\Omega$ is written as
\begin{equation}
\Omega =
 (1- {x_1})^{\delta_1}
 (1- {x_2})^{\delta_2}
=
  (1-\bar {x}_1)^{\bar {\delta}_1}
  (1-\bar {x}_2)^{\bar {\delta}_2}.
\label{eq:r52}
\end{equation}
The equation shows a correlation  between the fraction $x_i$
and the fractal dimension $\delta_i$.
(The scale transformation can be chosen so that
${\bar {\delta}_1}={\bar {\delta}_2}$.)
Thus, the measure is an invariant under simultaneity of the
scale transformation of Lorenz invariants $x_i$ and
multiplicative transformation  of $\delta_i$.

Taking into account the scale transformation
of $x_i$ and $\delta_i$
similar to (\ref{eq:r52})
the measure  of the interaction $\Omega$ can be written as
\begin{equation}
\Omega = V^{\delta},
\label{eq:r51}
\end{equation}
where $\delta$ is the coefficient (fractal dimension) describing
the intrinsic structure of the interaction constituents
   revealed at high energies.
The factor $V$ is part of the full phase-space of fractions
$\{x_1,x_2\}$ corresponding to such parton-parton
collisions in which the inclusive particle can be produced.
The scaling variable $z$ can be written in the form
  \begin{equation}
   z=z_0\cdot V^{-\delta}
  \label{eq:r511}
  \end{equation}
where $z_0=\sqrt{{\hat s}_{\bot}}/\rho_A(s)$.
The variable $z$ has character of a fractal measure.
The fractal property of the collision reveals itself so that
only the part of all multiple scattering corresponding to the phase space
$V^{\delta}$ produces the inclusive particle.

{\section{Properties of $z$-scaling}}

In this section we discuss properties of the $z$-scaling for particles  ($\pi^{\pm,0}, K^{\pm}, \bar p,
\gamma, jet$) produced in $h-h$, $h-A$ and $A-A$ collisions. They are the energy  and angular independence of
data $z$-presentation, the power law of the scaling function at very high-$p_T$,  $A$- and $F$-dependence of
$z$-scaling. All properties are asymptotic ones because they reveal themselves at extreme conditions (high
$\sqrt s$ and $p_T$). Numerous experimental data on inclusive cross sections at high-$p_T$ obtained  at U70
\cite{Protvino,turch93}, ISR \cite{Angel}-\cite{Eggert},\cite{WA70}-\cite{R110},
\cite{AFS},\cite{ClarkHe}-\cite{Karabar},
 SpS
\cite{UA2g,UA6,UA1,UA2,Povlis,Albrecht,WA80,WA98}, and Tevatron
\cite{Cronin,fris83,CDFpho,D0pho,CDF,D0,E557,CDFj,D0A,Alverson,E706} were used in the analysis.

{\subsection{Energy independence}}

The energy independence of data $z$-presentation means that the scaling function $\psi(z)$ has the same shape
for different $\sqrt s$ over a wide $p_T$ range.


Figures 1(a)-4(a)  show the dependence  of the cross section of the $p-p$ and $\pi -p$ interactions on
transverse momentum $q_T$ at $\sqrt s=11-62~GeV$ and a produced angle of $90^0$. We would like to note that
the data cover a wide transverse momentum range, $q_T = 1-14~GeV/c$.

Some features of the hadron spectra should be stressed. The first one is the  strong  dependence of the cross
section on energy $\sqrt s$.
 The second feature is a tendency
 that  the difference between hadron yields increases
 with the transverse momentum and the energy $\sqrt s$.
 The third one is a non-exponential behavior of the spectra
 at $q_{T}>1~GeV/c$.

 Figures 1(b)-4(b) show $z$-presentation of the same data sets.
 Taking into account the experimental errors we can conclude that
 the scaling function $\psi(z)$ demonstrates energy
 independence over a wide energy and transverse momentum
 range at $\theta_{cm}^{h N} \simeq 90^0$.

 Figures 5-8 show the energy dependence of data $q_T$-
 and  $z$-presentation
 for direct-$\gamma$ and $jet$ production in $p-p$ and $\bar p-p$
 collisions. Experimental data on cross sections were obtained
 at ISR \cite{WA70}-\cite{R110}, \cite{AFS}, SpS \cite{UA2g,UA6} and Tevatron
\cite{CDFpho,D0pho,CDF,D0,E557,CDFj}.

 One can see that all data sets reveal
 the property of the energy independence of $\psi(z)$
 in $z$-presentation.

{\subsection{Angular independence}}

The angular independence of data $z$-presentation means that the scaling function $\psi(z)$ has the same shape
for different values of an angle $\theta $ of produced particle over a wide $p_T$ range and $\sqrt s$. Taking
into account the energy independence of $\psi(z)$ it will be enough to verify the property at some $\sqrt s$.

To analyze the angular dependence of the scaling function $\psi(z)$ we use some data sets. The first one
obtained at ISR \cite{Lloyd,Akes1} includes the results of  measurements of the invariant cross section
$Ed^3\sigma/dq^3$  at  $\sqrt s = 23~GeV$ over a momentum and angular ranges of $q_T=1.2-3.~GeV/c$ and
$\theta_{cm}^{pp}=15^0-90^0$, respectively.  A strong dependence of the cross section on the angle of the
produced $\pi^0$-meson  was experimentally found. The second one is the D0 data \cite{D0} for direct
$\gamma$'s produced in $\bar p-p$ collisions at $\sqrt s =1800~GeV$ and the rapidity range $\eta=0.0-2.5$. The
D0 data on invariant cross sections of jets production in $\bar p-p$ collisions at $\sqrt s =1800~GeV$ and
$\eta = 0.-3.$ are taken from \cite{D0A}.

Figures 9(a)-11(a) show the dependence  of the cross section of $\pi^0$-meson, direct-$\gamma$ and $jet$
production in $p-p$  and $\bar p-p$ collisions  on transverse momentum  at fixed  $\sqrt s$ and for different
rapidity intervals.

Figures 9(b)-11(b) demonstrate $z$-presentation of the same data sets. The obtained results show that the
function $\psi(z)$ is independent of the angle $\theta$ over a wide range  at ISR and Tevatron energies. This
is the experimental confirmation of the angular scaling of data $z$-presentation.

{\subsection{Power law}}

Here, we discuss a new feature of data $z$-presentation
for $\pi^0$-meson, direct-$\gamma$ and $jet$ production.
This is the power law of the scaling function, $\psi(z) \sim z^{-\beta}$.

As seen from  Figures 2(b),5(b)-8(b), 10(b) and 11(b) the data sets demonstrate a linear $z$-dependence of
$\psi(z)$ on the log-log scale at high $z$. The quantity  $\beta $ is the slope parameter.

Taking into account the accuracy of the available experimental data, we can conclude that the behavior of
$\psi(z)$ for $\pi^0$-mesons  produced in $p-p$ collisions reveals a power dependence and the value of the
slope parameter $\beta_{pp}^{pi}$ is independent of the energy $\sqrt s$ \cite{Rog1} over a wide range of high
transverse momentum. The mean values of the slope parameter for $\pi^0$-meson production in $p-p$ and $\bar
p-p$ \cite{Banner} were found to be 7.30 and 5.75, respectively.

The values of the slope parameter for direct $\gamma$ and $jet$ production  in $p-p$ and $\bar p-p$ collisions
were found to be different ones so that $\beta_{pp} > \beta_{\bar pp}$.

 The mean values of
 $\beta_{pp}^{\gamma}$ and $\beta_{\bar pp}^{\gamma}$
are found to be 5.91 and  5.48, respectively. Direct photons are mainly produced in $p-p$  and $\bar p-p$
collisions through the Compton  and annihilation processes, respectively. This fact can be the main reason of
different values of the slope parameters
 $\beta_{pp}^{\gamma}$ and $\beta_{\bar pp}^{\gamma}$.

The values of the slope parameter found for jet and dijet  production in $\bar p-p$ and jet production in
$p-p$ collision differ considerably each other. It give us possibility to study the features of these
different processes in the same approach. The mean values of $\beta $  for jet and dijet  production in $\bar
p-p$ were found to be  $5.3 \pm 0.2$  and $4.75 \pm 0.05$, respectively. The mean value of $\beta_{pp}^{jet} $
found from the $p-p$ data \cite{AFS,E557} is equal to $5.92 \pm 0.17$. Single jets are mainly produced in
softer environment than dijets. The fact can be responsible for the different values of the slope parameters
$\beta_{\bar pp}^{jet}$ and  $\beta_{\bar pp}^{2jet}$. Note that the value of $\beta_{\bar pp}^{jet}$ is
constant with the  high accuracy and independent of rapidity intervals and the energy $\sqrt s$ for the
separate data sets (see Ref.\cite{Dedovich}).

 Thus we can conclude based on the obtained results  that  behavior of $\psi(z)$
      for $\pi^0$-meson, direct-$\gamma $  and  $jet$ production
      at high $z$
      reveals the power dependence with high accuracy.
      The value of the slope
      parameter is independent of the colliding energy
      $\sqrt s$ and  the angle of produced particles  over a wide
      range of high transverse momentum.
 The existence of the power law, $\psi(z) \sim z^{-\beta }$,
 means, from our point of view, that
 the mechanism of particle formation reveals fractal behavior.

{\subsection{A-dependence}}

A study of  $A$-dependence of particle production in $h-A$ and $A-A$ collisions is traditionally connected
with  nuclear matter influence  on particle  formation. The difference  between  the cross sections of
particle production on free and bound nucleons is normally  considered as an indication of unusual physics
phenomena like EMC-effect
$J/\psi$-suppression
and Cronin effect \cite{Cronin}.

A-dependence of $z$-scaling for particle production in $p-A$ collisions was studied in \cite{Z01}. It was
established $z$-scaling for every nuclei ($A=D-Pb$) and type of produced particles ($\pi^{\pm,0}, K^{\pm},
\bar p$). The symmetry transformation
 of the scaling function $\psi(z)$ and variable $z$ under
the scale transformation $z\rightarrow \alpha_A z$, $\psi \rightarrow \alpha_A^{-1} \psi $ was suggested to
compare the scaling function for different nuclei. It was found that $\alpha$ depends on the atomic number
only and can be parameterized by the formula $\alpha(A)=0.9A^{0.15}$
\cite{Z01}.

Figures 12(a) and 13(a) demonstrate  the spectra of $\pi^+$- and $\pi^0$-mesons produced in proton-nucleus
collisions. The $z$-presentations of the same data are shown in Figures 12(b) and 13(b).

Similar results was obtained for particle production in $\pi^--A$ collisions too \cite{Tok01}. The dependence
of the inclusive cross section of $K^-$-mesons on transverse momentum and the corresponding  scaling function
versus  $z$ at $p_{lab}=200$ and $300~GeV/c$ are shown in Figures 14(a) and 14(b), respectively.

The interesting results are obtained under the comparison  of the scaling function of direct-$\gamma$
production in $p-p$  and  $p-Be$ collisions. The experimental data on the inclusive cross section as a
function of transverse momentum are shown in Figure 15(a). The shape of the scaling functions (Figure 15(b))
is found to be a linear one on log-log scale for both cases. The value of the slope parameter
$\beta_{pBe}^{\gamma}$ is equal to  5.97 and $\beta_{pBe}^{\gamma}\simeq \beta_{pp}^{\gamma}$. The fact means
that the nuclear matter changes the probability of photon formation with different formation length  $z$ and
do not change  the fractal dimension of the mechanism of photon formation (photon "dressing"). The nuclear
effect can be described by the ratio
 $R^{A/p}(z,A)$ of the scaling functions of nucleus and proton.
One can see from Figure 15(b) that the ratio is practically independent of $z$. The function $R^{A/p}$ can
depend on atomic number only. The experimental verification of the statement should be performed. Taking into
account an experimental accuracy of data used in the analysis, the obtained results show that the fractal
dimension  $\delta$ and  the slope parameter $\beta$ is independent of A. Direct $\gamma$ is considered to be
a good probe to study the Quark Gluon Plasma (QGP) formation. Therefore the experimental investigations of
$A$-dependence of $z$-scaling for direct photons  produced in hadron-nucleus collisions at RHIC and LHC
energies are very important to obtain any indications on nuclear phase transition.

It is assumed that high energy-density nucleon matter produced in heavy-ion collisions could give an direct
indication of phase transition to new state of nuclear matter, QGP. High-$p_T$ $\pi^0$-meson spectra should be
sensitive to the transition \cite{Wang1}. Recently, the WA80 \cite{WA80}  and WA98 \cite{WA98} Collaborations
have measured the $\pi^0$-meson spectra  of $S-S$, $S-Au$ and $Pb-Pb$ collisions at $p_{lab} = 200~AGeV/c$ and
$158~AGeV/c$, respectively. Therefore, we  compare $p_T$-dependence of the cross sections for proton-nucleus
and nucleus-nucleus collisions using the available experimental data \cite{Albrecht,WA80,WA98}. In our
analysis, we use the data for high-$p_T$ $\pi^0$ production in light-ion ($d-d$, $\alpha-\alpha$) collisions
obtained at ISR \cite{ClarkHe,Angelis,Karabar}. The cross sections were measured for $p_T>2~GeV/c$ at an angle
of about $90^0$ and  $\sqrt s = 26$ and $31~GeV$.

A change of the shape of the $p_T$ spectra  could evidence a  modification of the mechanism of particle
formation in nuclear matter.

Figure 16 shows dependence of inclusive cross section of $\pi^0$-meson produced in nucleus-nucleus collisions
on transverse momentum and the results of $z$-presentation of the same data sets. We found that the dependence
$\alpha$  on the atomic weight for $AA$ collisions can be parameterized by the  formula
$\alpha(A)=0.93A^{0.32}$  \cite{Rog2}. In the case  the fractal dimension $\delta$ is equal to 0.5. Thus the
available experimental data \cite{Povlis}-\cite{E706}, \cite{WA80}-\cite{Karabar}
give no strong indications of the scaling violation
over the range $z=4-20.$

New data on high-$p_T$ cross sections of $\pi^0$-mesons produced in $Au-Au$ collisions at RHIC are presented
by PHENIX Collaboration  in \cite{Gabor}. The  pion spectra for different centralities are shown in Figure 18
(a). We compared the scaling function  for minimum bias  data set with  function  found at ISR and SpS
energies. The obtained results are shown in Figure 18(b).  Note that the dramatic change of the value of the
fractal dimension $\delta $ was found. The scaling function at ISR and SpS energies corresponds to
$\delta=0.5$ whereas  at RHIC energies the value of $\delta$ is found to be 3.5. This is indication that
nuclear environment changes essentially the mechanism of particle formation. The increase of fractal dimension
$\delta$ means that mechanism of  multiple scattering play an important role of particle formation in nuclear
medium created at RHIC energies. We suggest to use the change of the fractal dimension ("$\delta$-jump") as
signature of nuclear matter transition. Therefore it is of interest to investigate the energy dependence of
fractal dimension $\delta (s)$ and determine the shape of the dependence.

{\subsection{F-dependence}}

 As we mention above the physics meaning of the scaling function $\psi(z)$
is the probability density to produce particle with formation length $z$. The existence of the scaling is the
confirmation of self-similarity at different scales, regulated by the energy $\sqrt s$ and transverse momentum
$p_T$. The power law,  $\psi(z)\sim z^{-\beta}$, observed at very high-$p_T$ range is  characterized by the
slope parameter $\beta$ . It was found the independence of the parameter on $\sqrt s$  over a wide transverse
momentum range. Therefore it is of interest to study the dependence of the slope parameter $\beta$ on type of
produced particle ($\pi^{\pm,0}, K^{\pm}, \bar p$). The dependence is named as the $F(lavor)$-dependence of
$z$-scaling.

Figure 17(a) shows the scaling functions for produced particles with different flavor content. The symmetry
transformation of the scaling function $\psi \rightarrow \alpha_F^{-1} \psi $ and the variable  $z\rightarrow
\alpha_F z$ was used to compare the scaling function for different particles. The results give us the
indication on the existence of the universal asymptotic of $\psi(z)$ at high $z$ for different type of
particles, $\pi^{\pm,0}, K^{\pm}, \bar p$. The property of data $z$-presentation reflects new feature of
particle formation, the flavor independence of the scaling function in the asymptotic region. The verification
of the property  for other particles, $J/\psi, \Upsilon, D, B, W^{\pm}$ and $Z^0 $, will be possible at RHIC
and LHC and it is  important for understanding of the mechanism of particle formation.

{\section{$z-p_T$  plot}}

To know the kinematic region where the $z$-scaling  can be violated is very important for modelling of the
experiment. The $z-p_T$ plot allows us to determine the high transverse momentum range interesting for such
investigations.

Figure 17(b) shows the $z-p_{T}$ plot for $p-Be$ interaction  at $\sqrt s = 31-500~GeV$. As seen from Figure
17(a) the scaling function $\psi(z)$ is measured up to $z\simeq 20$. Therefore the kinematic ranges
interesting for searching new physics phenomena will be  $p_T > 8, 12, 20$ and $24~GeV/c$ at $\sqrt s = 31,
63, 200$ and $500~GeV$, respectively.

{\section{Conclusions}}

 Analysis of numerous experimental data on high-$p_T$ particle production
 in hadron-hadron, hadron-nucleus and nucleus-nucleus collisions
 obtained at ISR, SpS and Tevatron in the framework of $z$-scaling concept
 was presented. The general scheme of data $z$-presentation for different
 processes was formulated.

  The scaling function $\psi(z)$ and scaling variable $z$
  are expressed via the experimental quantities, the invariant
  inclusive cross section
  $Ed^3\sigma/dq^3$  and the multiplicity  density of charged
  particles $\rho_A(s,\eta)$.
 The physics interpretation the scaling function $\psi$ as a
 probability density to produce a particle with the formation length $z$
 is argued. The quantity  $z$ has the property of the fractal measure
 and $\delta $  is the fractal dimension describing the
 intrinsic structure of the interaction constituents revealed
 at high energies. The fractal dimensions of nucleon $\delta_N$, pion $\delta_{\pi}$
 and nuclei $\delta_A$ were found. They satisfy the relation $\delta_{\pi}<\delta_N << \delta_A$.

 It was shown that the properties of $z$-scaling, the energy and angular independence,
 the power law  $\psi(z)\sim z^{-\beta}$, $A$- and $F$-dependence are confirmed by
 the numerous experimental data obtained by different Collaborations at ISR, SpS and Tevatron.

 The new data on $\pi^0$-meson spectra in $Au-Au$ collisions
 obtained at RHIC were analyzed.  The  dramatic change of
 the fractal dimension ("$\delta$-jump") was found. The results is considered as one of the
 confirmation on creation of new state of nuclear matter.

Thus, the obtained results show that data $z$-presentation demonstrates general properties of the particle
production mechanism such as self-similarity, locality, scale-relativity and fractality. The properties reveal
themselves both in $h-h$, $h-A$ and $A-A$ collisions and  reflect the features of elementary constituent
formation.

The violation of $z$-scaling due to the change of the value of the fractal dimension $\delta$
 is suggested to
search for new  physics phenomena such as quark compositeness, new type of interactions, nuclear phase
transition,  fractal structure of space-time
 in hadron-hadron, hadron-nucleus and nucleus-nucleus collisions
at SpS, RHIC, Tevatron, HERA and LHC.

\vskip 1cm

{\Large \bf Acknowledgments} \vskip 0.5cm

The results presented in the report  are obtained with I.Zborovsky, Yu.Panebratsev, G.Skoro, E.Potrebennikova,
T.Dedovich and O.Rogachevski. The author would like to thank them for collaboration and many useful and
stimulating  discussions of the problem.

\vskip 1cm

{\small

\newpage
\begin{minipage}{4cm}

\end{minipage}

\vskip 4cm
\begin{center}
\hspace*{-2.5cm}
\parbox{5cm}{\epsfxsize=5.cm\epsfysize=5.cm\epsfbox[95 95 400 400]
{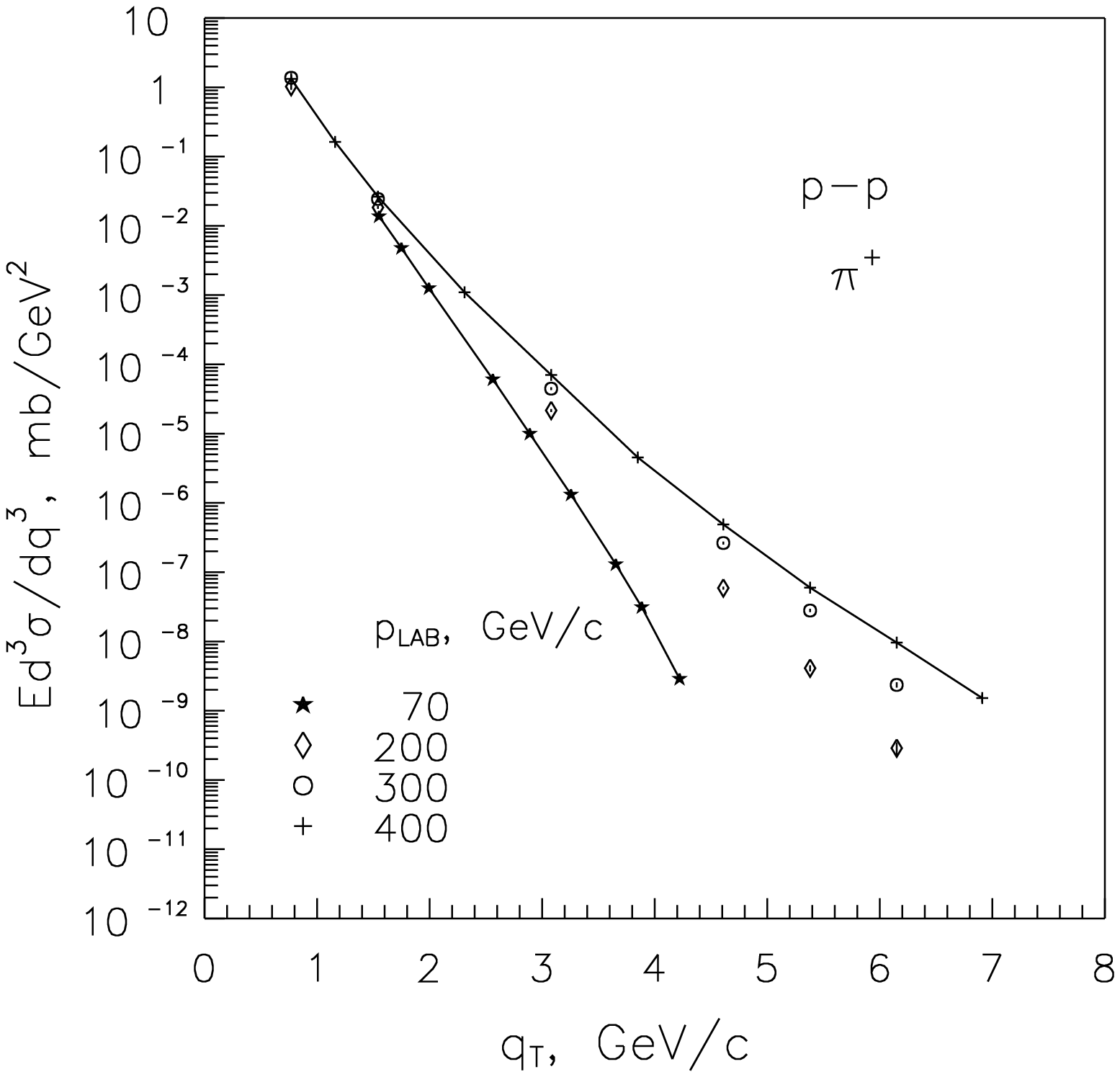}{}}
\hspace*{3cm}
\parbox{5cm}{\epsfxsize=5.cm\epsfysize=5.cm\epsfbox[95 95 400 400]
{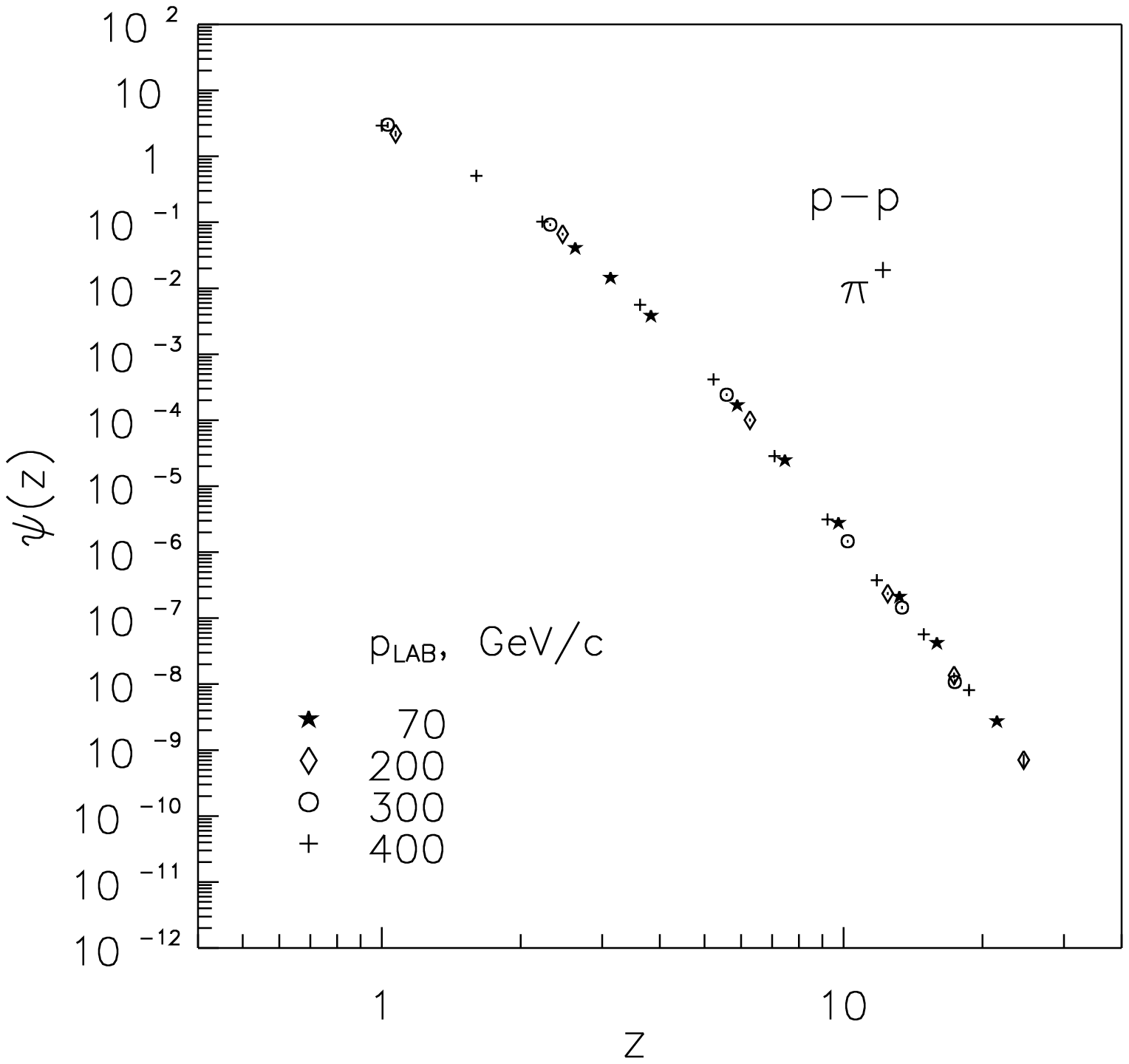}{}}
\vskip -0.5cm
\hspace*{0.cm} a) \hspace*{8.cm} b)\\[0.5cm]
\end{center}

{\bf Figure 1.} (a) The  inclusive differential cross sections for the $\pi^+$-mesons produced in $p-p$
collisions at $p_{lab} = 70, 200, 300$ and $ 400~GeV/c$ and $\theta_{cm}^{pp} \simeq 90^{0}$ as functions of
the transverse momentum  $q_{T}$. (b) The corresponding scaling function $\psi(z)$. Solid lines are obtained
by fitting  of the data at  $p_{lab} = 70$, and $400$. Experimental data are taken from
\cite{Cronin,Protvino}.

\vskip 5cm

\begin{center}
\hspace*{-2.5cm}
\parbox{5cm}{\epsfxsize=5.cm\epsfysize=5.cm\epsfbox[95 95 400 400]
{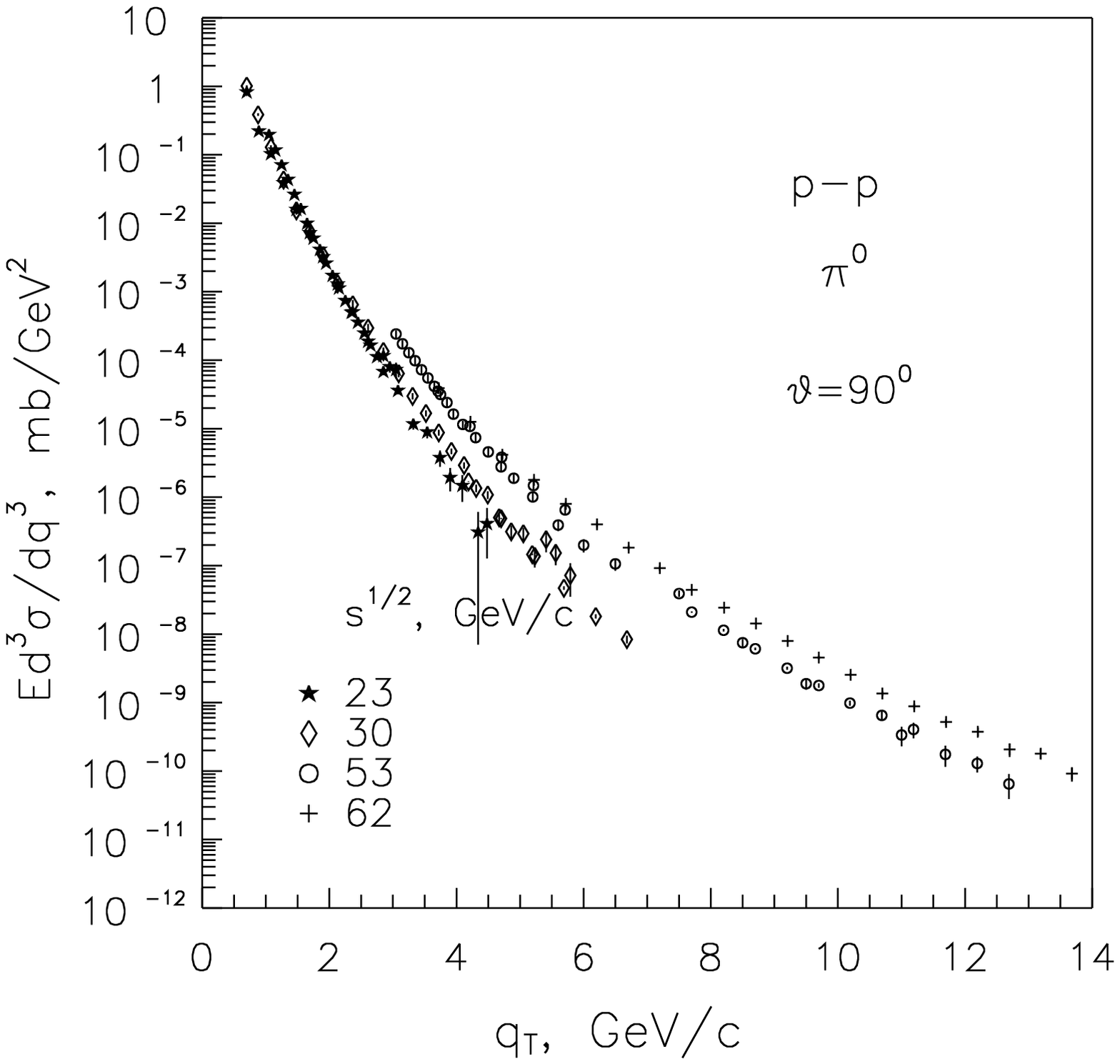}{}}
\hspace*{3cm}
\parbox{5cm}{\epsfxsize=5.cm\epsfysize=5.cm\epsfbox[95 95 400 400]
{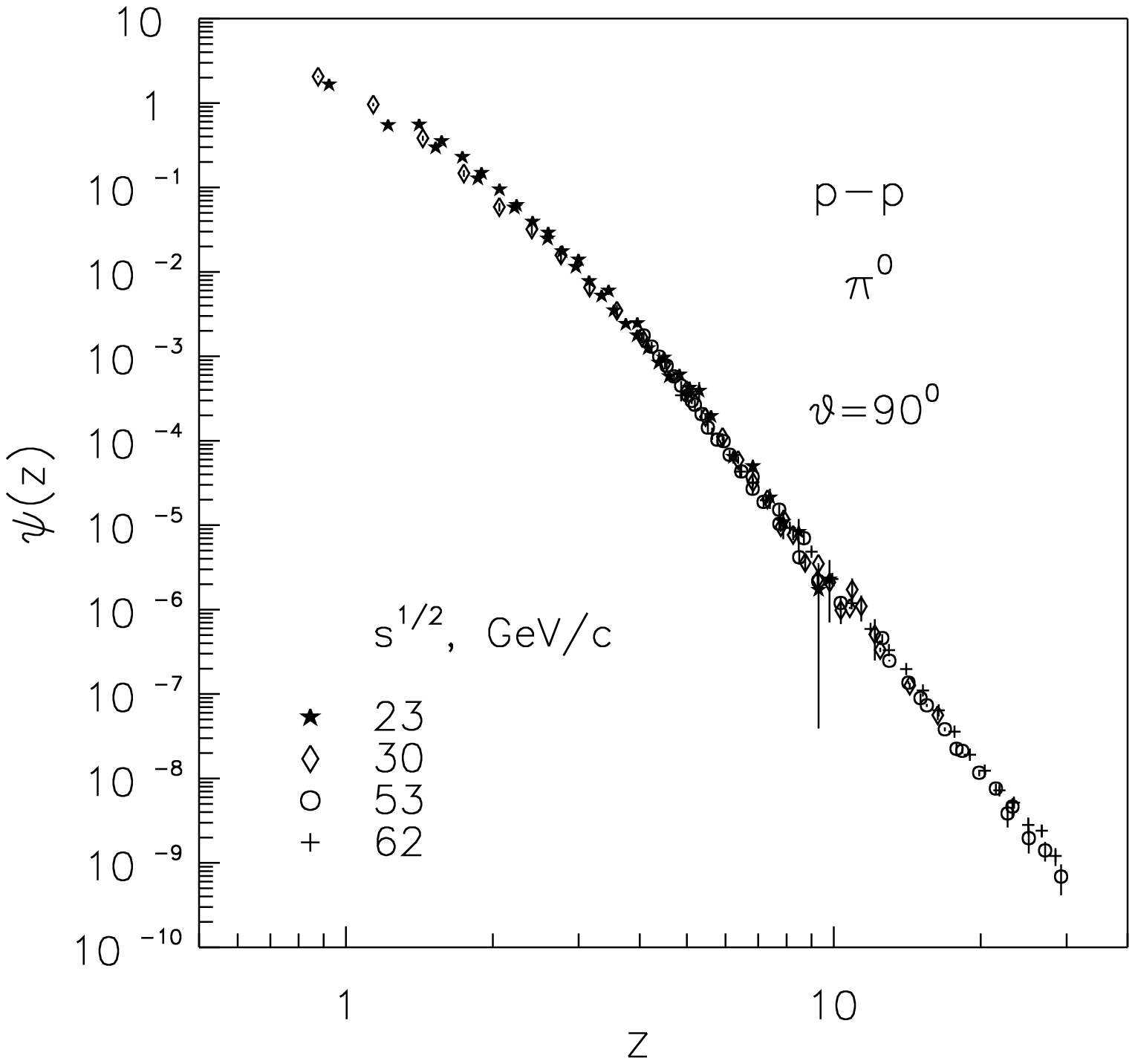}{}}
\vskip -1.cm
\hspace*{0.cm} a) \hspace*{8.cm} b)\\[0.5cm]
\end{center}

{\bf Figure 2.} (a) The dependence of  the inclusive cross section of $\pi^0$-meson production on transverse
momentum $q_{\bot}$ in $pp$  collisions at $\sqrt s = 23,30,53$ and 62$~GeV$ and an angle $\theta_{cm}^{pp}$
of $90^0$. The experimental data on the cross section are taken from \cite{Angel,Kou1,Kou3,Lloyd,Eggert}. (b)
The corresponding scaling function $\psi(z)$.

\newpage
\begin{minipage}{4cm}

\end{minipage}

\vskip 4cm
\begin{center}
\hspace*{-2.5cm}
\parbox{5cm}{\epsfxsize=5.cm\epsfysize=5.cm\epsfbox[95 95 400 400]
{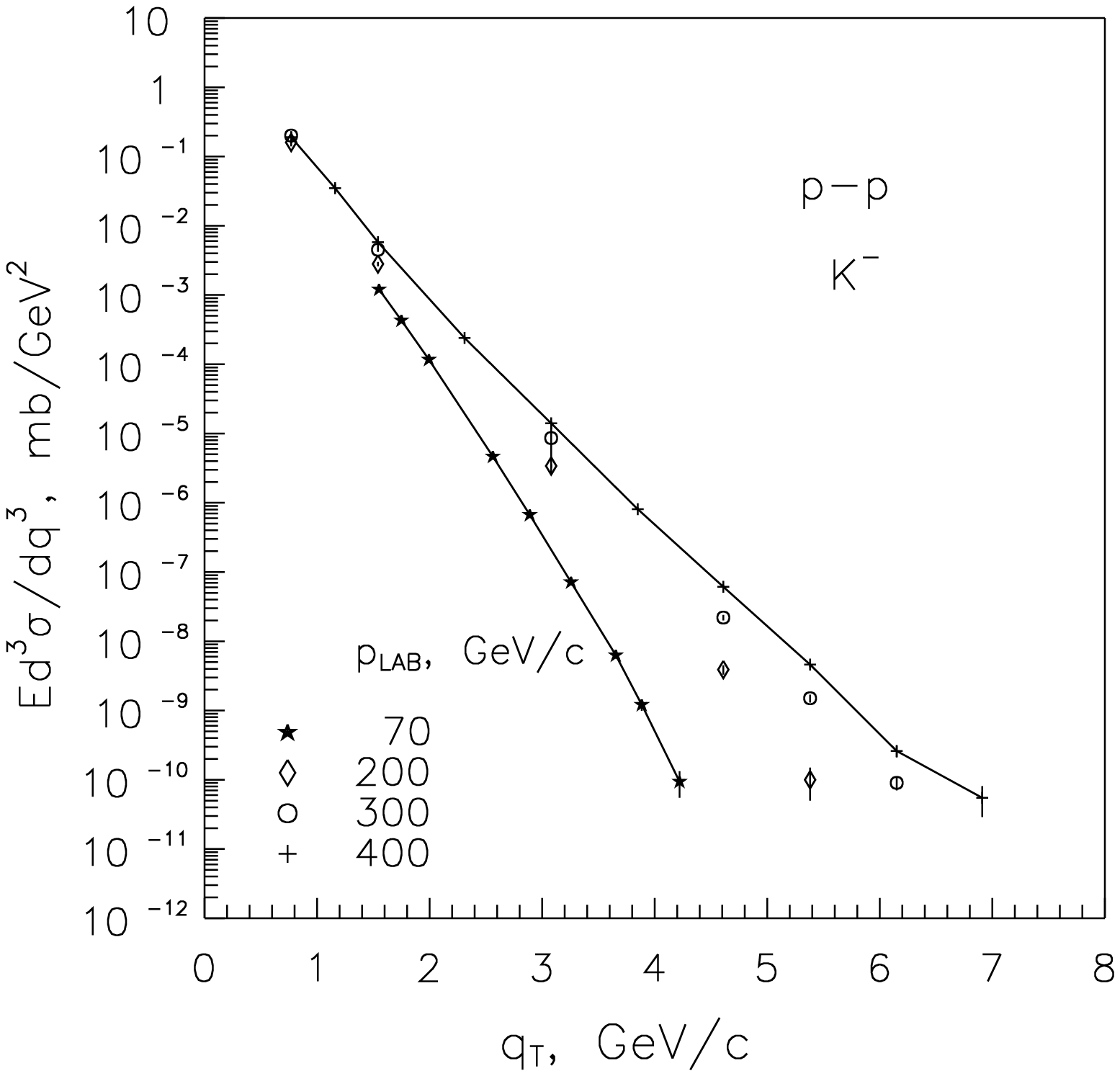}{}}
\hspace*{3cm}
\parbox{5cm}{\epsfxsize=5.cm\epsfysize=5.cm\epsfbox[95 95 400 400]
{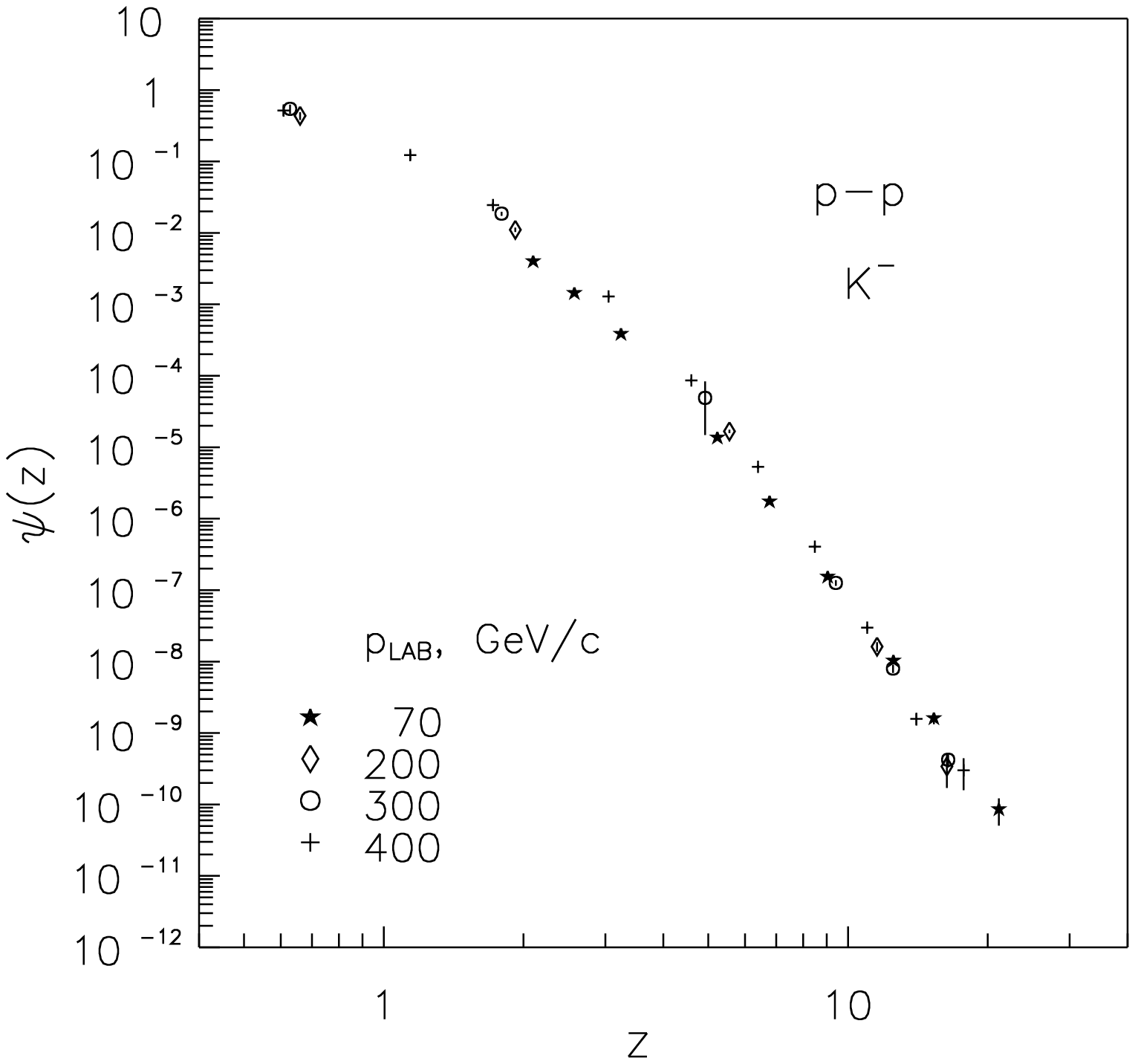}{}}
 \vskip -0.5cm
\hspace*{0.cm} a) \hspace*{8.cm} b)\\[0.5cm]
\end{center}

{\bf Figure 3.} (a) The inclusive differential cross sections for  $K^-$-mesons produced in $p-p$ collisions
at $p_{lab} = 70, 200, 300$ and $ 400~GeV/c$ and $\theta_{cm}^{pp} \simeq 90^{0}$ as functions of the
transverse momentum  $q_{T}$.  Solid lines are obtained by fitting  of the data at  $p_{lab} = 70$, and
$400~GeV/c$.  Experimental data are taken from \cite{Cronin,Protvino}. (b) The corresponding scaling function
$\psi(z)$.

\vskip 5cm

\begin{center}
\hspace*{-2.5cm}
\parbox{5cm}{\epsfxsize=5.cm\epsfysize=5.cm\epsfbox[95 95 400 400]
{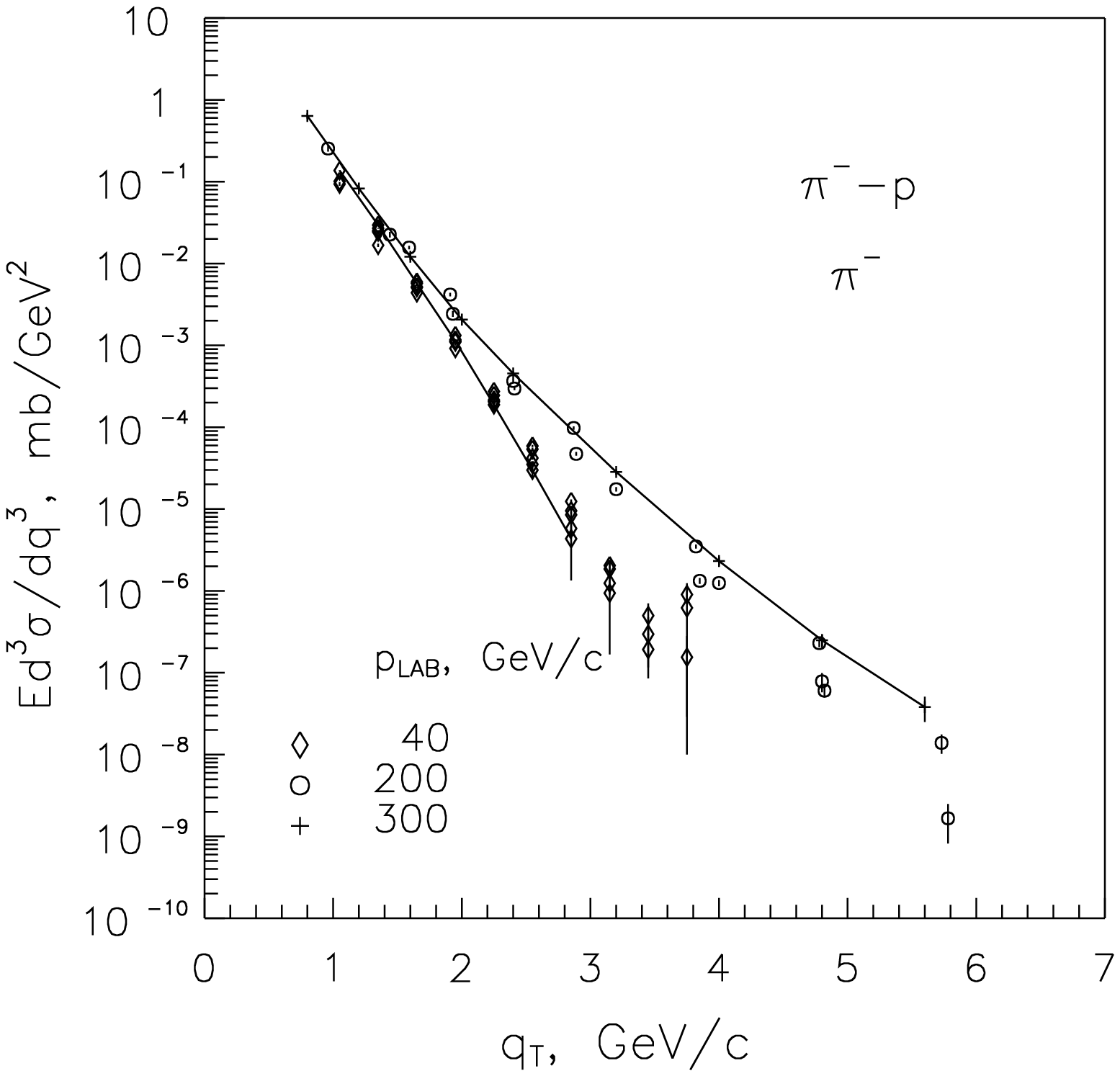}{}}
\hspace*{3cm}
\parbox{5cm}{\epsfxsize=5.cm\epsfysize=5.cm\epsfbox[95 95 400 400]
{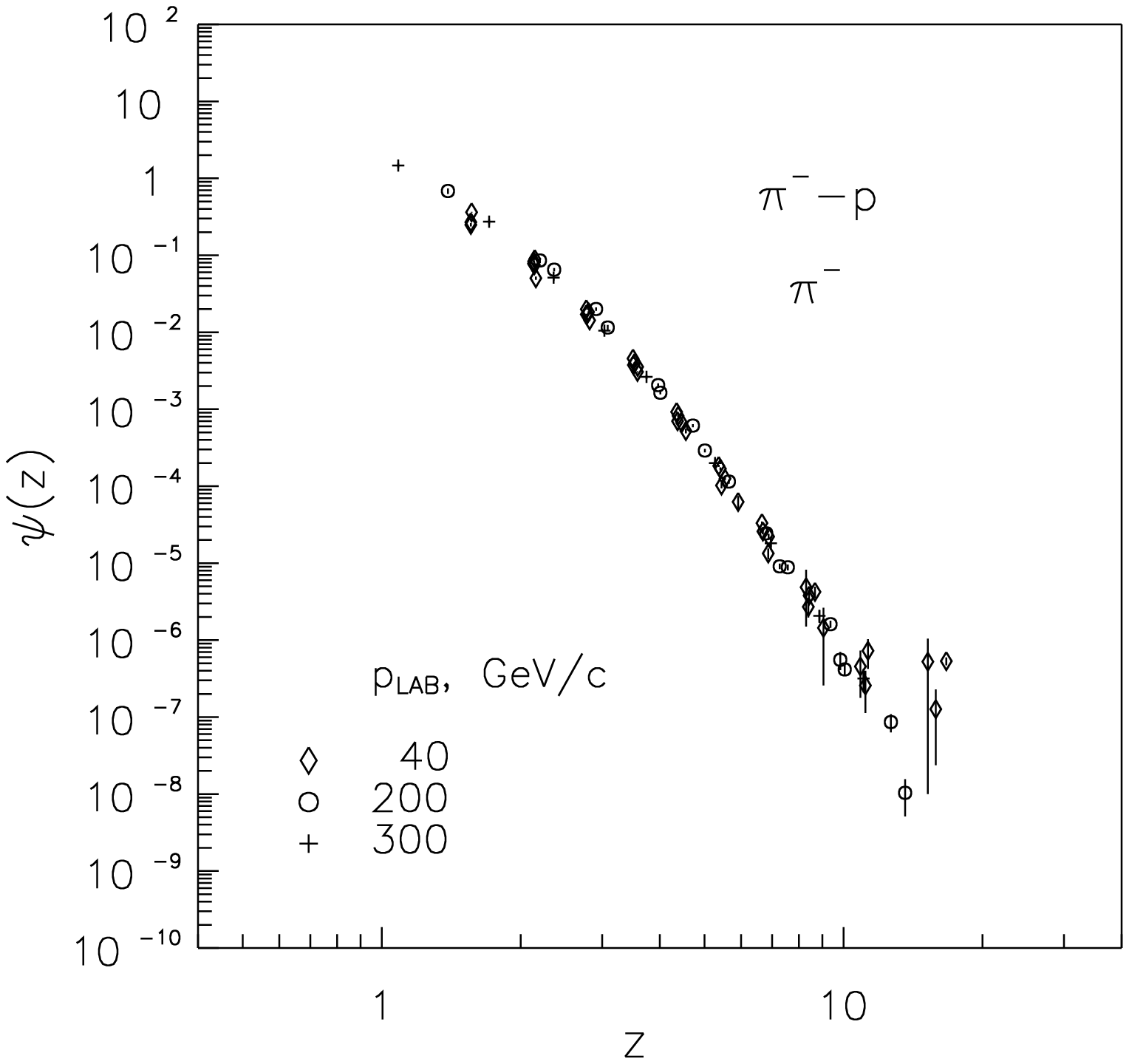}{}}
\vskip -1.cm
\hspace*{0.cm} a) \hspace*{8.cm} b)\\[0.5cm]
\end{center}

{\bf Figure 4.}
 (a) Dependence of  the
 inclusive cross section of $\pi^-$-meson  production
 on transverse momentum $q_{T}$ at $p_{lab} = 40, 200$ and $300~GeV/c$
 and $\theta_{cm}^{\pi p} \simeq 90^{0}$
 in  $\pi^--p$ collisions.
 Experimental data are taken from
 \cite{fris83,turch93}.
 (b) The corresponding scaling function $\psi(z)$.

\newpage
\begin{minipage}{4cm}

\end{minipage}

\vskip 4cm
\begin{center}
\hspace*{-2.5cm}
\parbox{5cm}{\epsfxsize=5.cm\epsfysize=5.cm\epsfbox[95 95 400 400]
{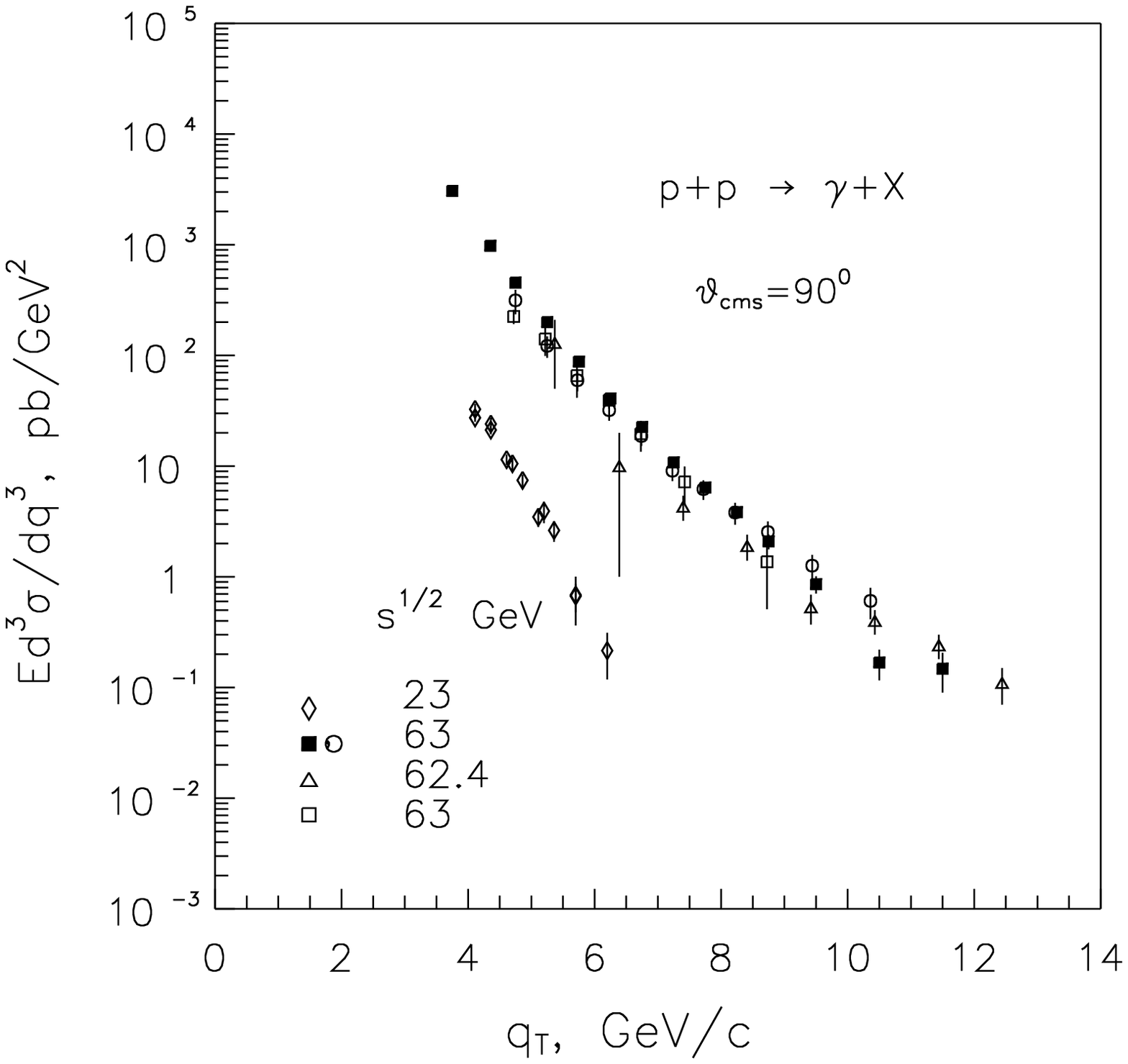}{}}
\hspace*{3cm}
\parbox{5cm}{\epsfxsize=5.cm\epsfysize=5.cm\epsfbox[95 95 400 400]
{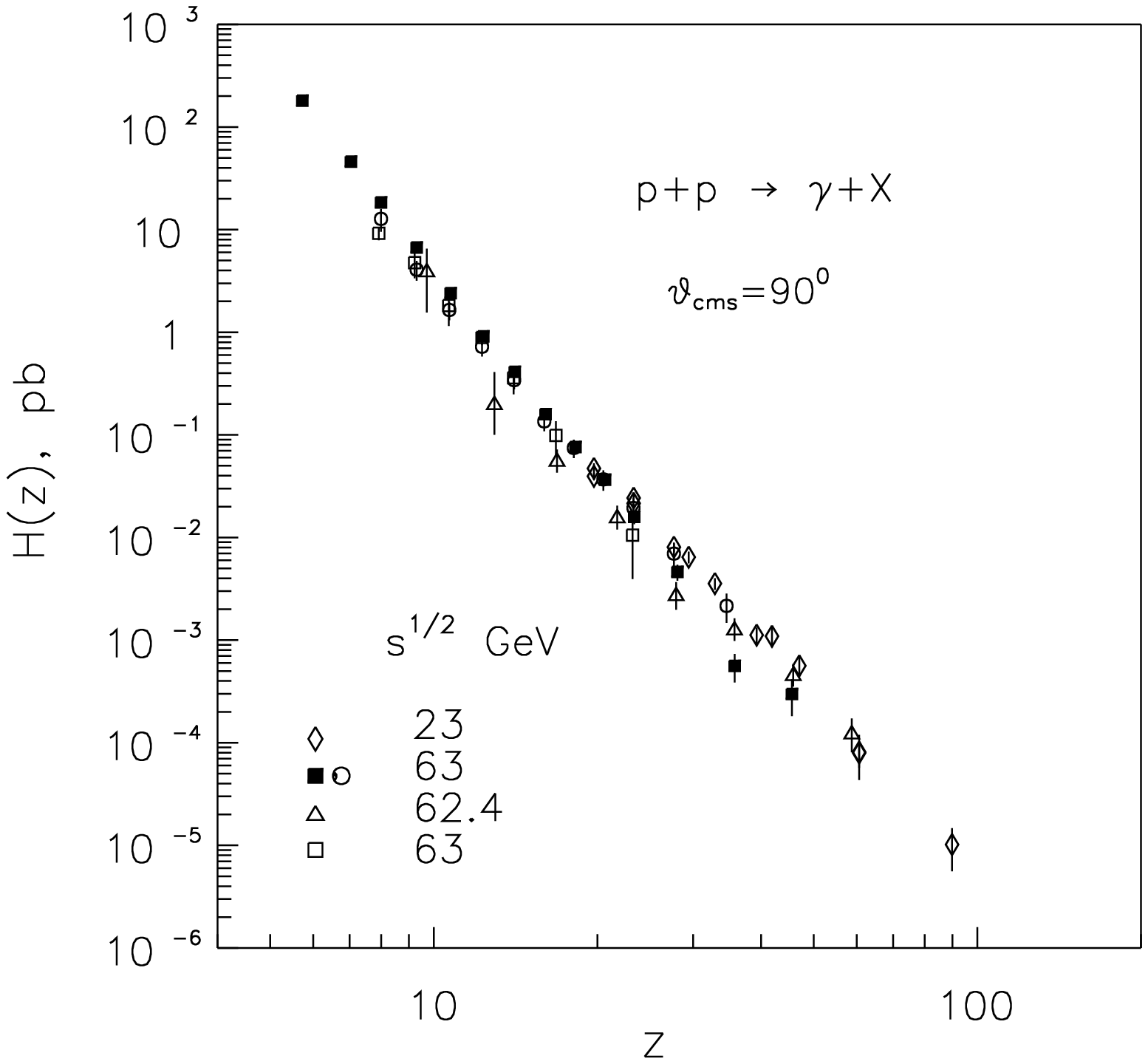}{}}
\vskip -0.5cm
\hspace*{0.cm} a) \hspace*{8.cm} b)\\[0.5cm]
\end{center}

{\bf Figure 5.} (a) Dependence of the inclusive cross section of direct photon production in  $pp$ collisions
on $q_T$ at energy $\sqrt s= 23$ and $63~GeV$ and  pseudorapidity $\eta\simeq 0$. Experimental data on the
cross section $\diamond$ - WA70 \cite{WA70}, $\bullet$ - R806 \cite{R806}, $\circ$ - R807 \cite{R807},
$\triangle$ - R108 \cite{R108}, $\overline{\sqcup}$ - R110 \cite{R110} are used. (b) The corresponding scaling
function $H(z)$.

\vskip 5cm

\begin{center}
\hspace*{-2.5cm}
\parbox{5cm}{\epsfxsize=5.cm\epsfysize=5.cm\epsfbox[95 95 400 400]
{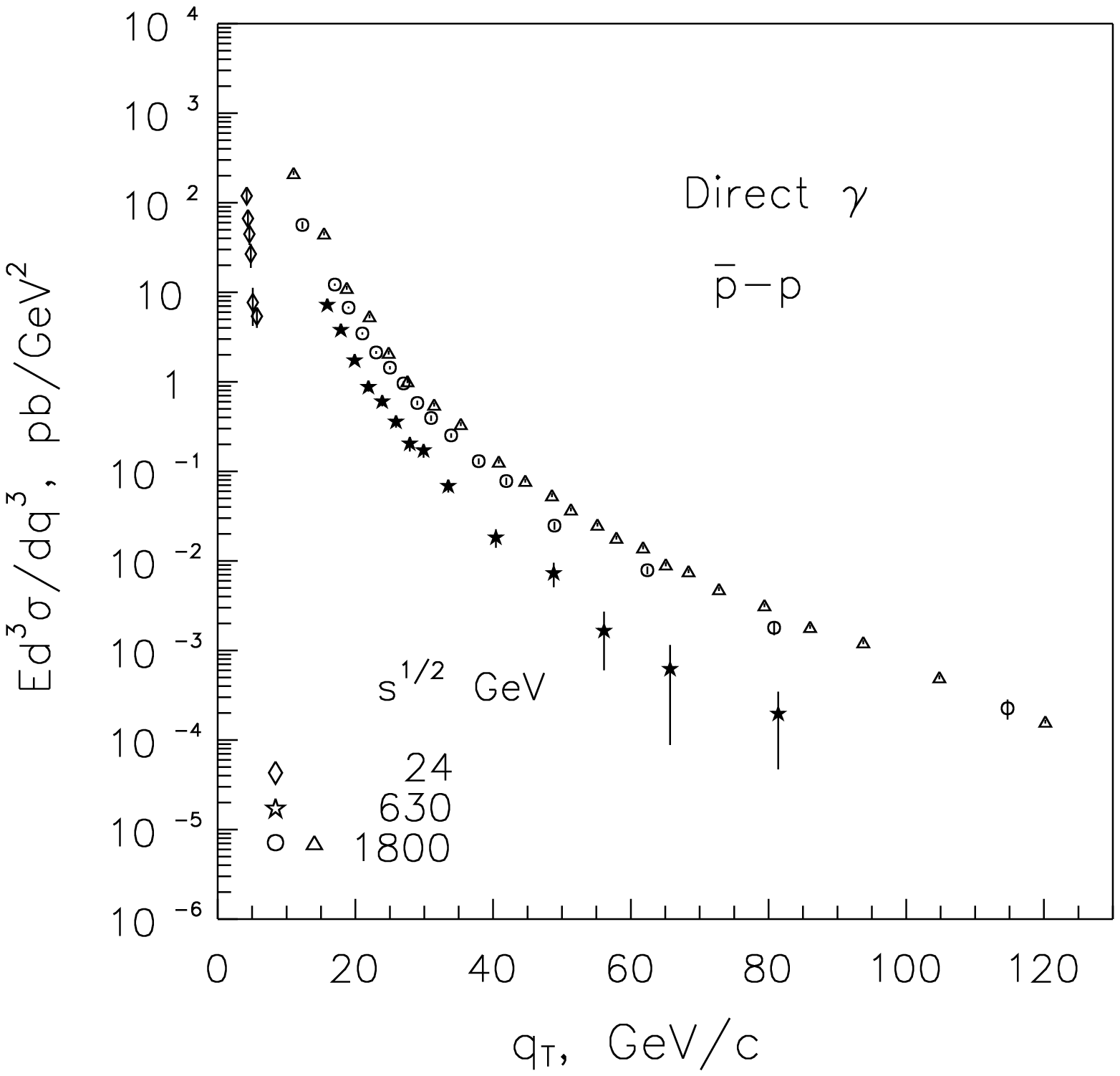}{}}
\hspace*{3cm}
\parbox{5cm}{\epsfxsize=5.cm\epsfysize=5.cm\epsfbox[95 95 400 400]
{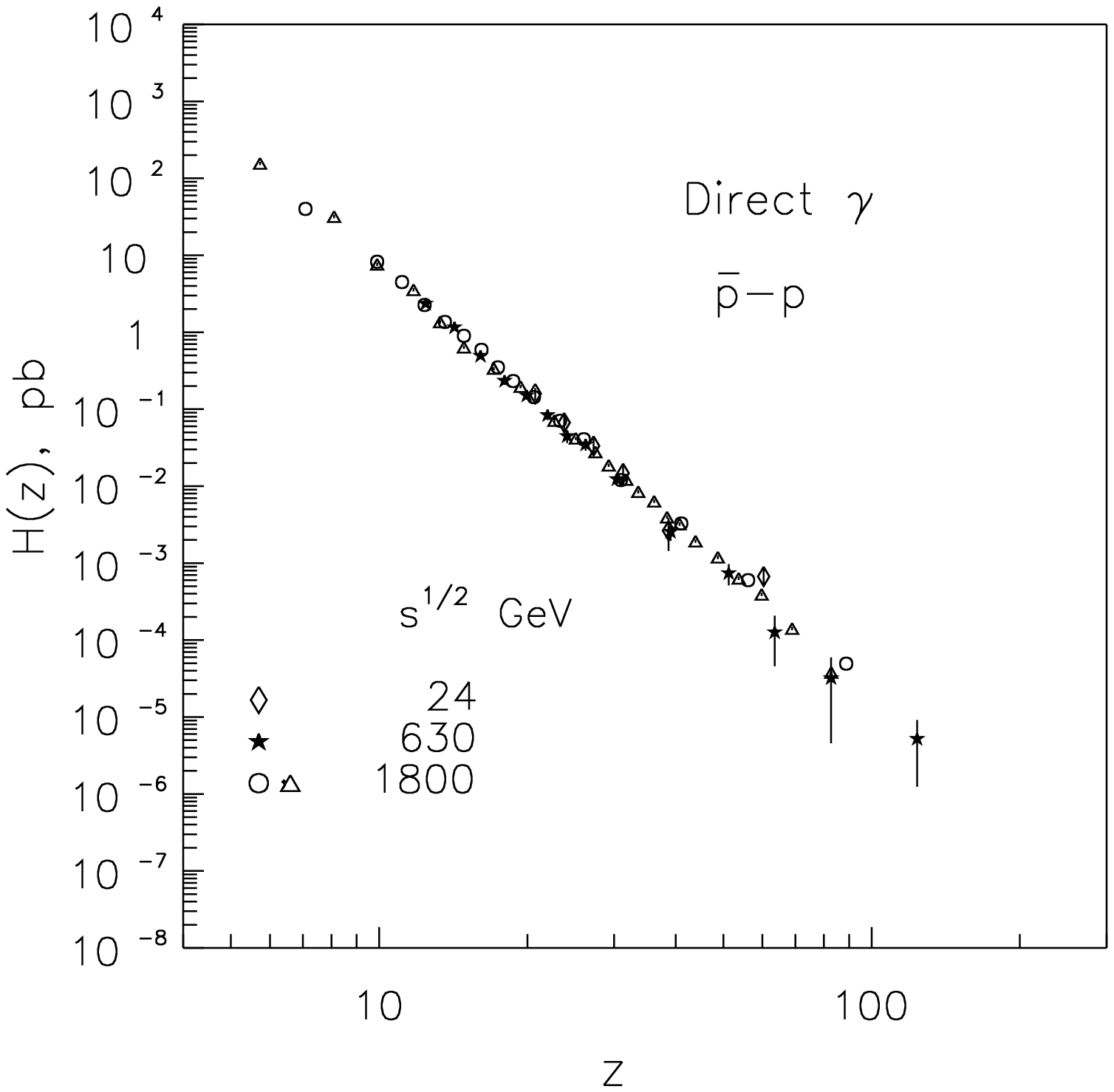}{}}
\vskip -1.cm
\hspace*{0.cm} a) \hspace*{8.cm} b)\\[0.5cm]
\end{center}

{\bf Figure 6.} (a) Inclusive differential cross section for prompt $\gamma$-production in $\bar pp$
collisions as a function of transverse momentum $q_T$ for different c.m.s. energies $\sqrt s =24, 630$ and $
1800~GeV$ and at a produced angle $\theta = 90^0$. Experimental data are taken from $\star$ - UA2 \cite{UA2g},
$\diamond$ - UA6 \cite{UA6}, $\circ $ - CDF \cite{CDFpho} and $\triangle $ - D0 \cite{D0pho}. (b)
Corresponding scaling function $H(z)$.

\newpage
\begin{minipage}{4cm}

\end{minipage}

\vskip 4cm
\begin{center}
\hspace*{-2.5cm}
\parbox{5cm}{\epsfxsize=5.cm\epsfysize=5.cm\epsfbox[95 95 400 400]
{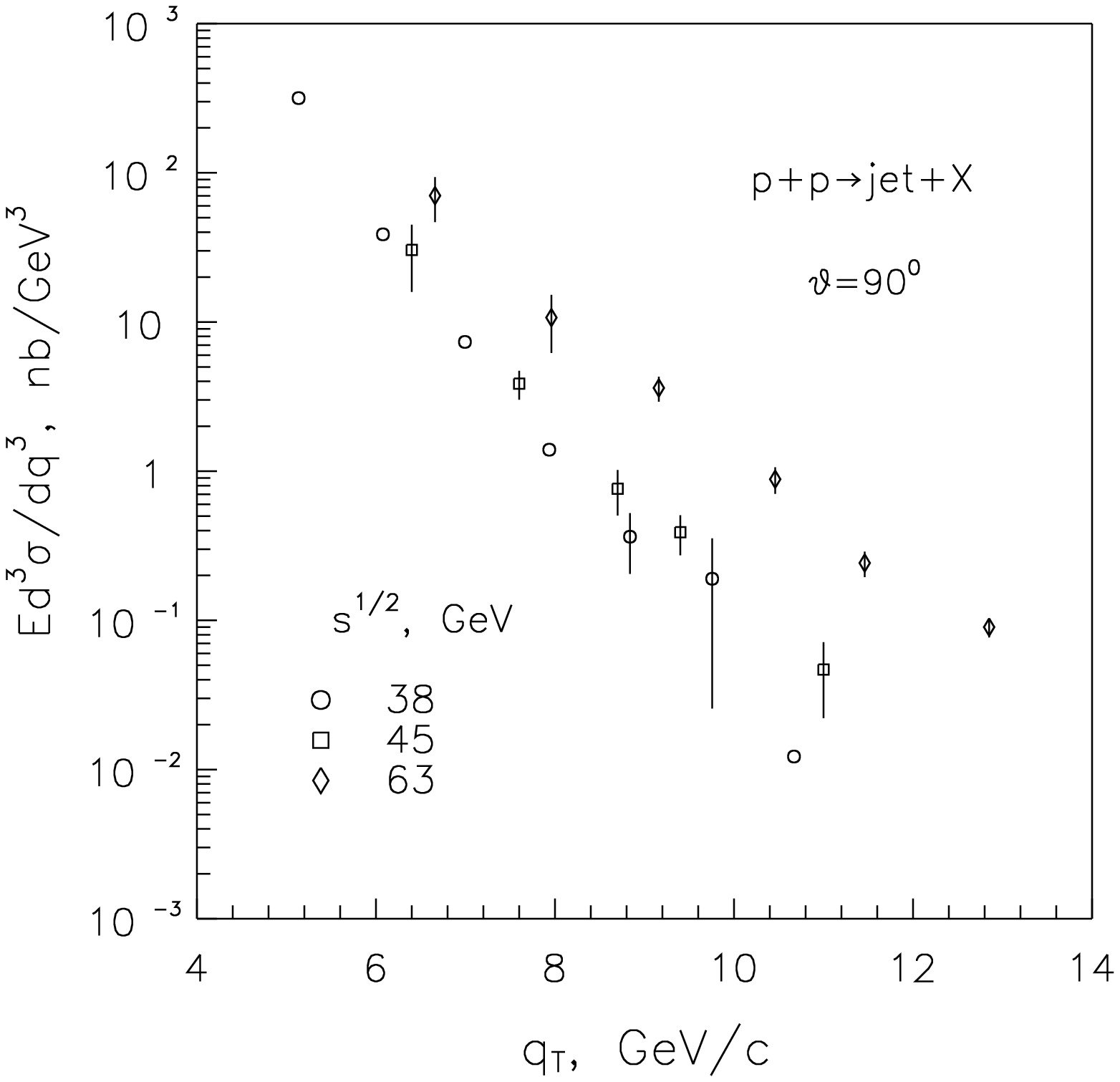}{}}
\hspace*{3cm}
\parbox{5cm}{\epsfxsize=5.cm\epsfysize=5.cm\epsfbox[95 95 400 400]
{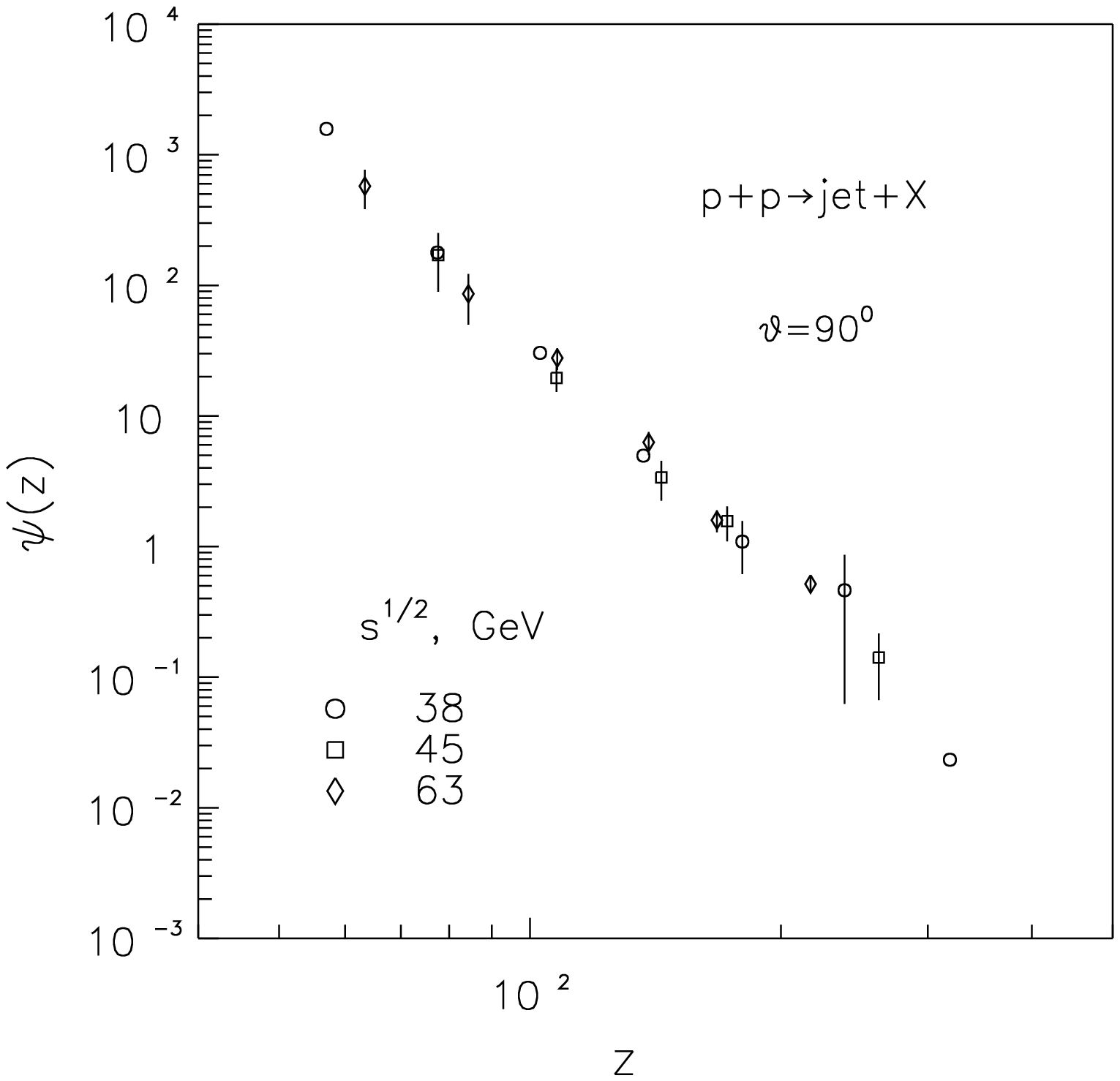}{}}
\vskip -0.5cm
\hspace*{0.cm} a) \hspace*{8.cm} b)\\[0.5cm]
\end{center}

{\bf Figure 7.} (a) The dependence of the inclusive cross section of jet production in $pp$ collisions at
$\sqrt s  = 38.8,45$ and $63~GeV/c$ and central rapidity range on transverse momentum $q_T$. Experimental data
of the cross sections obtained by the AFS and E557 Collaborations
 are taken from  \cite{AFS,E557}.
(b) The corresponding scaling function $\psi(z)$.

\vskip 5cm

\begin{center}
\hspace*{-2.5cm}
\parbox{5cm}{\epsfxsize=5.cm\epsfysize=5.cm\epsfbox[95 95 400 400]
{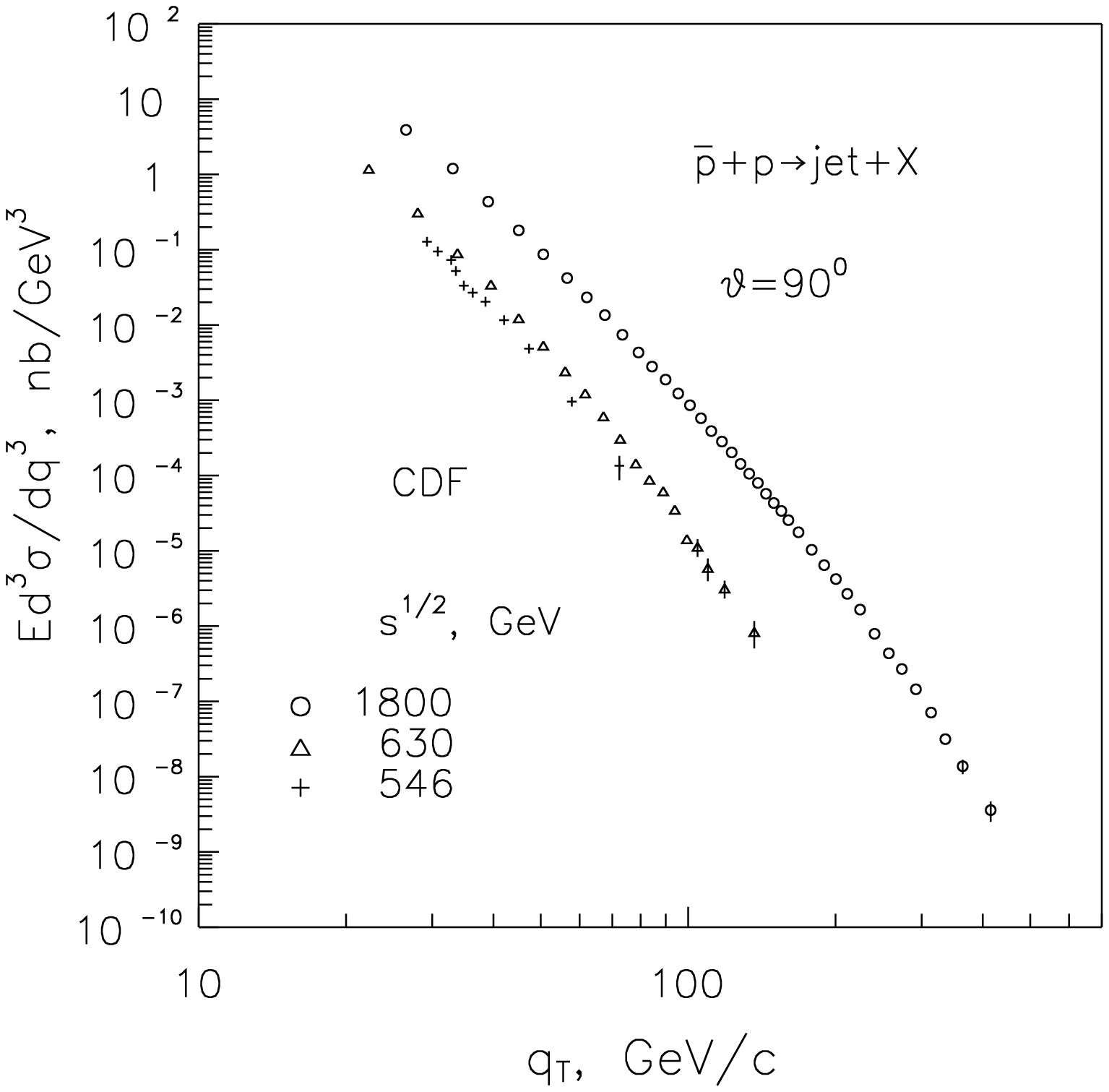}{}}
\hspace*{3cm}
\parbox{5cm}{\epsfxsize=5.cm\epsfysize=5.cm\epsfbox[95 95 400 400]
{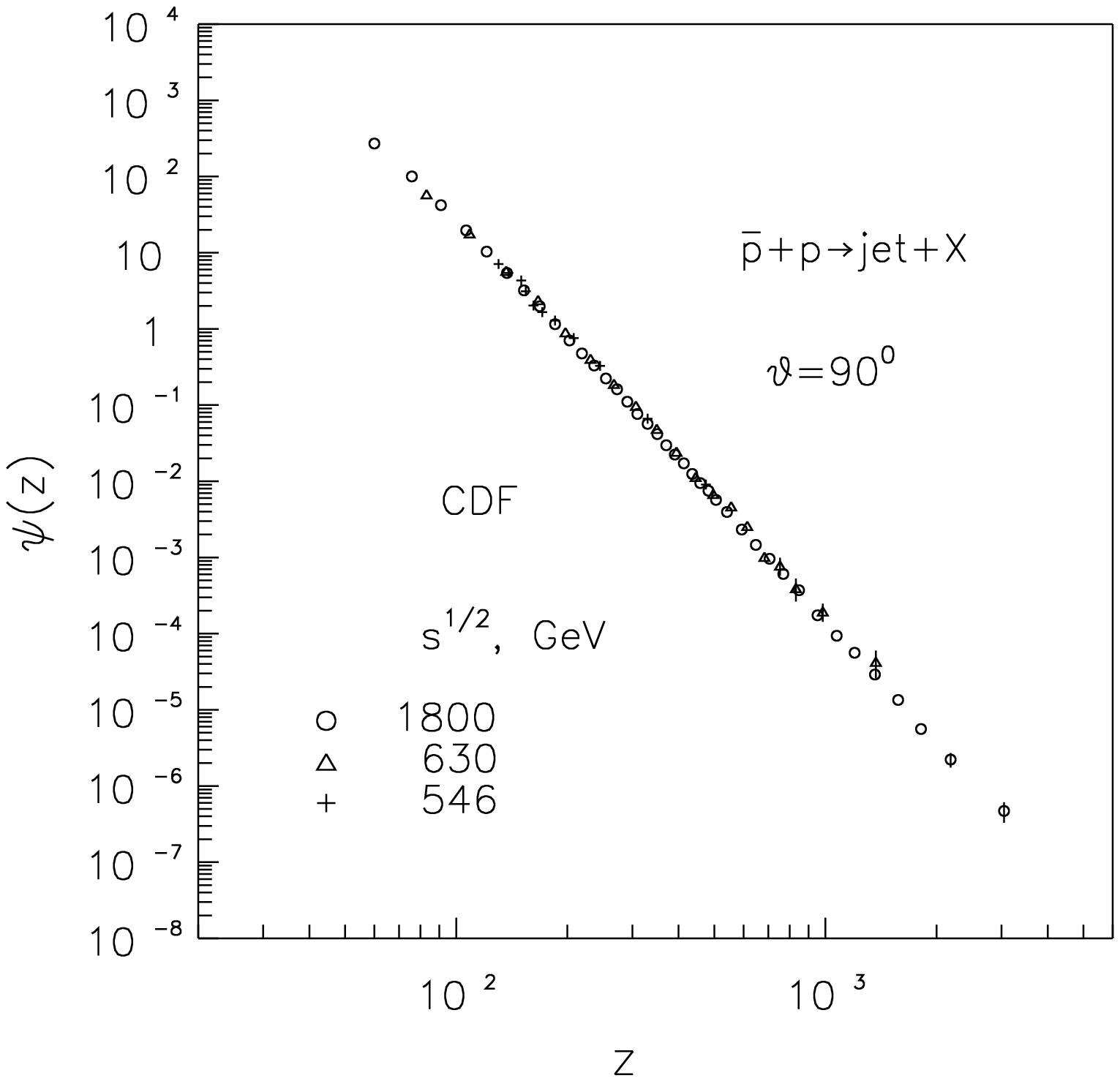}{}}
\vskip -1.cm
\hspace*{0.cm} a) \hspace*{8.cm} b)\\[0.5cm]
\end{center}

{\bf Figure 8.} (a) The dependence of the inclusive cross section of jet production in $\bar pp$ collisions at
$\sqrt s  = 546,630$ and $1800~GeV/c$ and central rapidity range on transverse momentum $q_T$. Experimental
data of the cross sections obtained by CDF Collaboration  are taken from  \cite{CDFj}. (b) The corresponding
scaling function $\psi(z)$.

\newpage
\begin{minipage}{4cm}

\end{minipage}

\vskip 4cm
\begin{center}
\hspace*{-2.5cm}
\parbox{5cm}{\epsfxsize=5.cm\epsfysize=5.cm\epsfbox[95 95 400 400]
{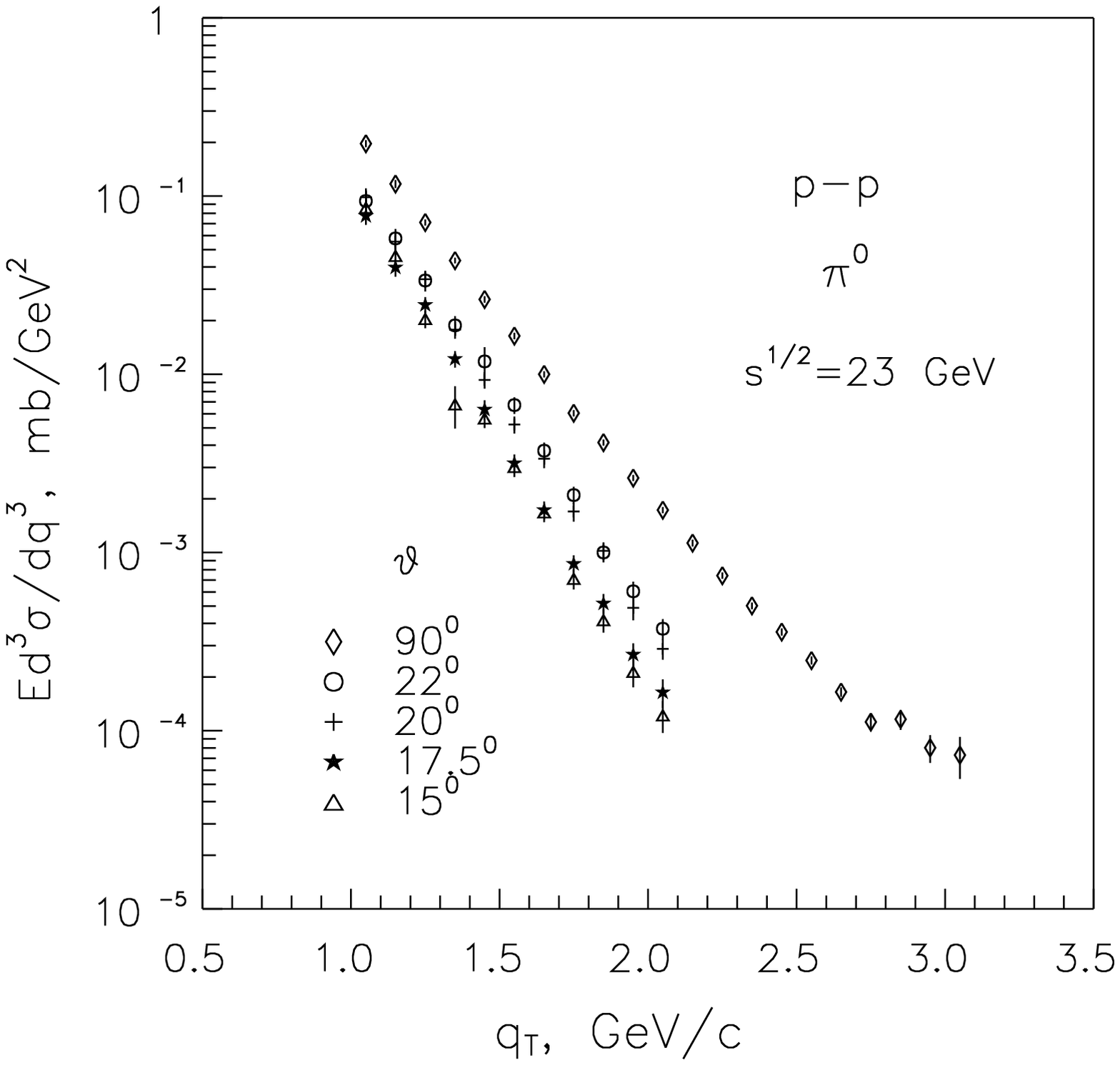}{}}
\hspace*{3cm}
\parbox{5cm}{\epsfxsize=5.cm\epsfysize=5.cm\epsfbox[95 95 400 400]
{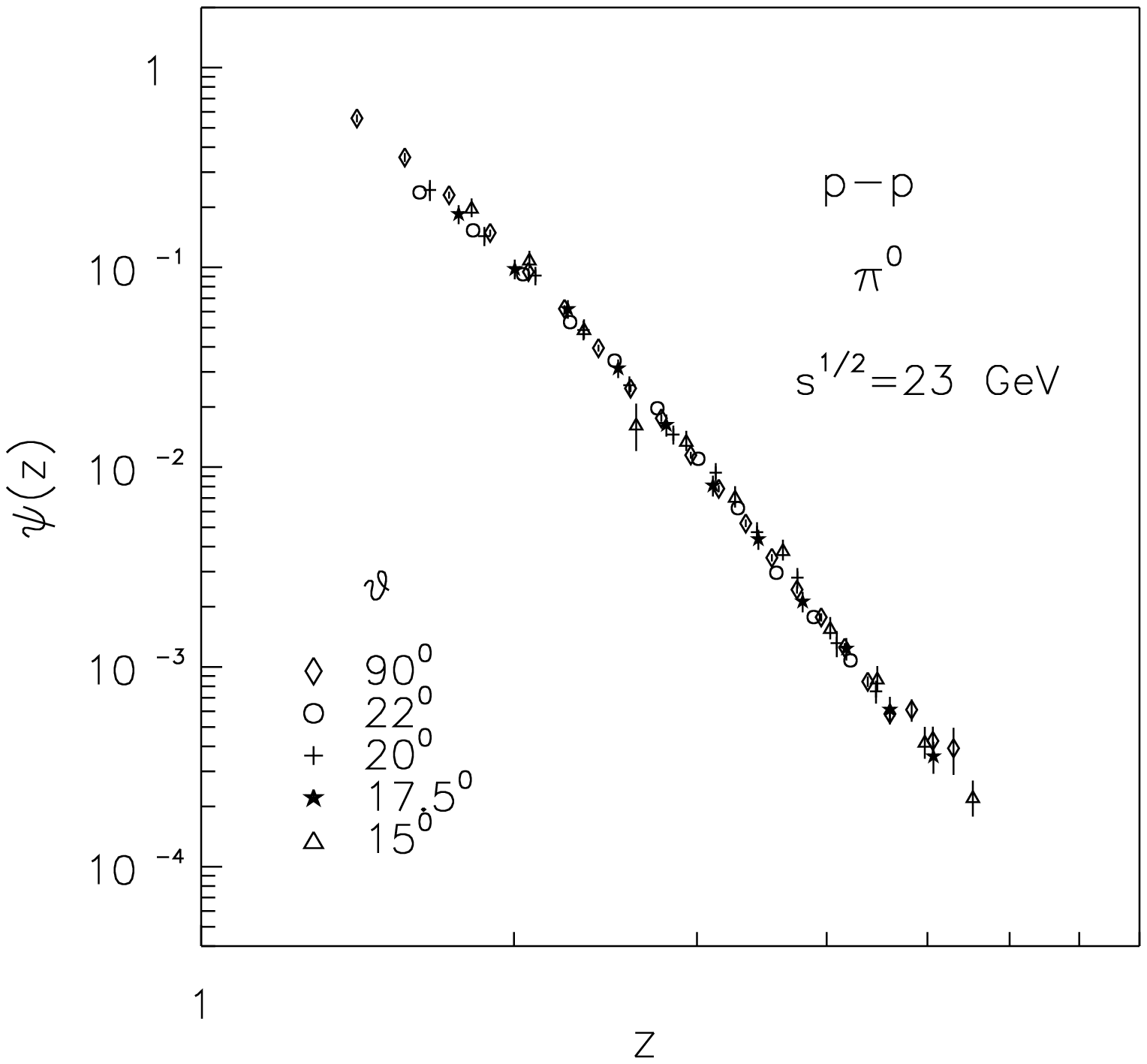}{}}
\vskip -0.5cm
\hspace*{0.cm} a) \hspace*{8.cm} b)\\[0.5cm]
\end{center}

{\bf Figure 9.} (a) The dependence of  the inclusive cross section of $\pi^0$-meson production in $p-p$
collisions at $\sqrt s = 23~GeV$ and an angle $\theta=15^0-90^0$ on transverse momentum $q_T$. The
experimental data on the cross sections are taken from \cite{Lloyd}. (b) The corresponding scaling function
$\psi(z)$.

\vskip 5cm

\begin{center}
\hspace*{-2.5cm}
\parbox{5cm}{\epsfxsize=5.cm\epsfysize=5.cm\epsfbox[95 95 400 400]
{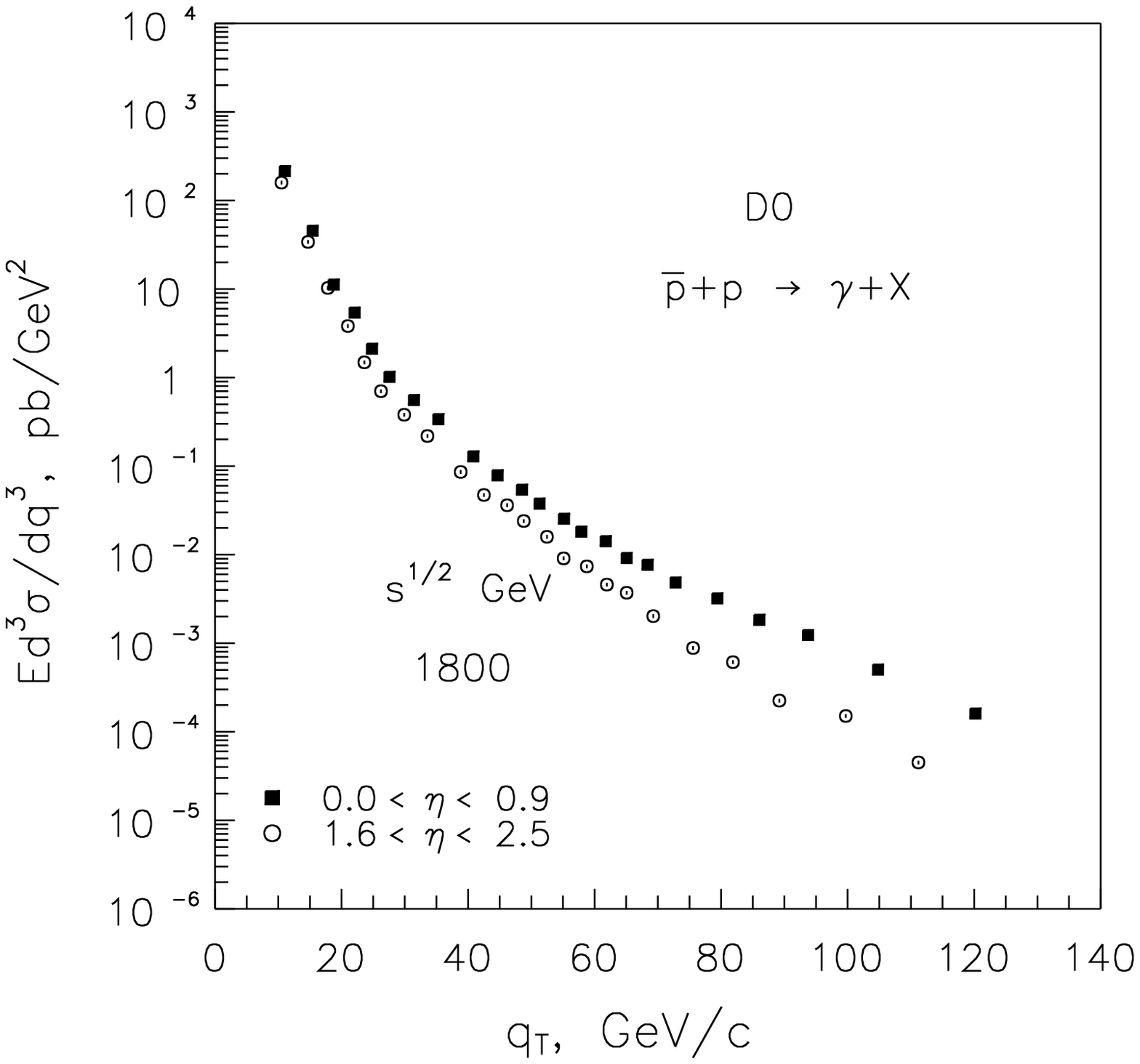}{}}
\hspace*{3cm}
\parbox{5cm}{\epsfxsize=5.cm\epsfysize=5.cm\epsfbox[95 95 400 400]
{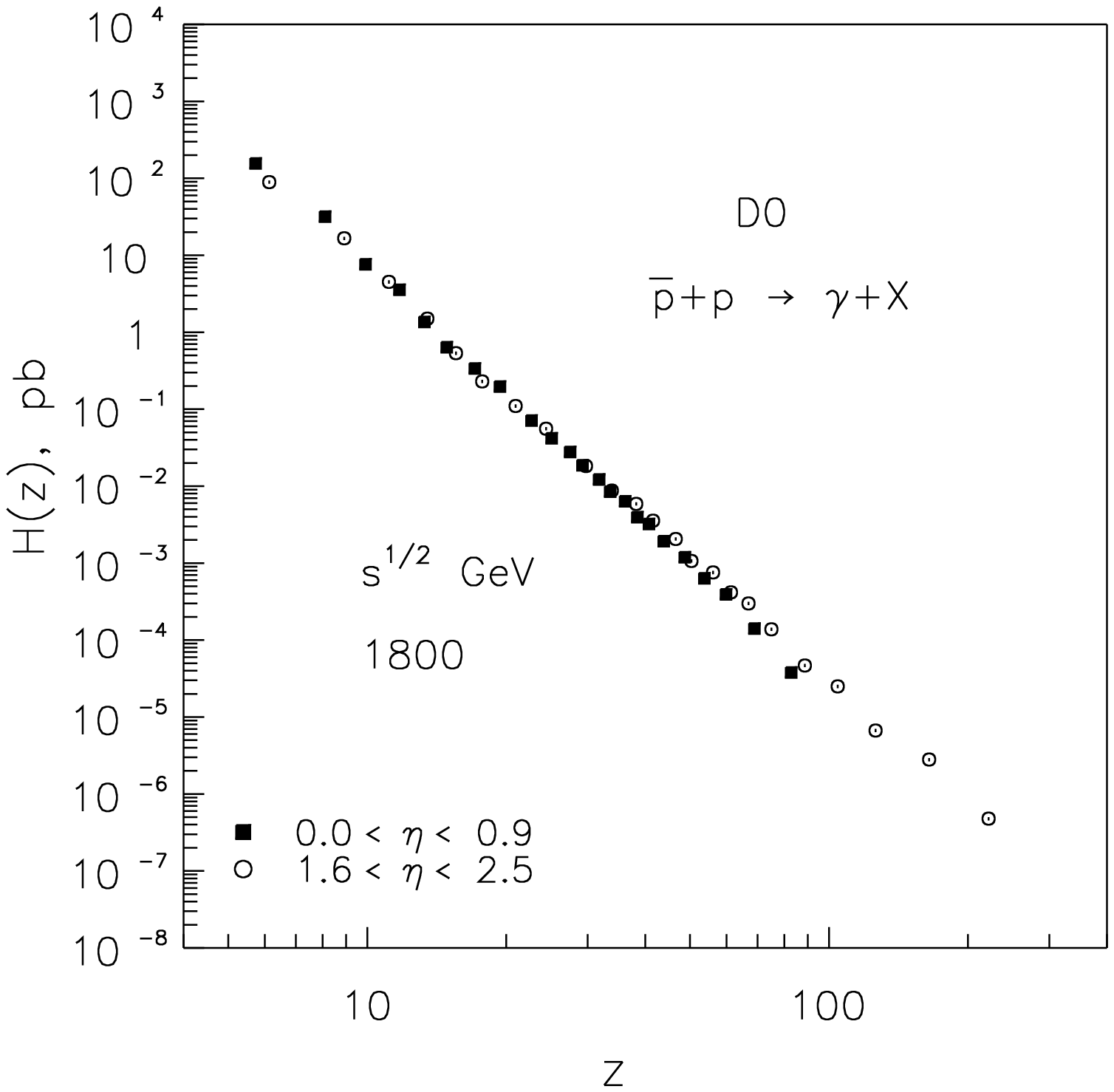}{}}
\vskip -1.cm
\hspace*{0.cm} a) \hspace*{8.cm} b)\\[0.5cm]
\end{center}

{\bf Figure 10.} (a) Dependence of  the inclusive cross section of  direct photon production in  $\overline
p-p$ collisions  on momentum $q_T$ for different pseudorapidity intervals at energy $\sqrt s= 1800~GeV$.
Experimental data on  the cross sections
 obtained by the D0 Collaboration  are taken from \cite{D0}.
(b) The corresponding  function $H(z)$.

\newpage
\begin{minipage}{4cm}

\end{minipage}

\vskip 4cm
\begin{center}
\hspace*{-2.5cm}
\parbox{5cm}{\epsfxsize=5.cm\epsfysize=5.cm\epsfbox[95 95 400 400]
{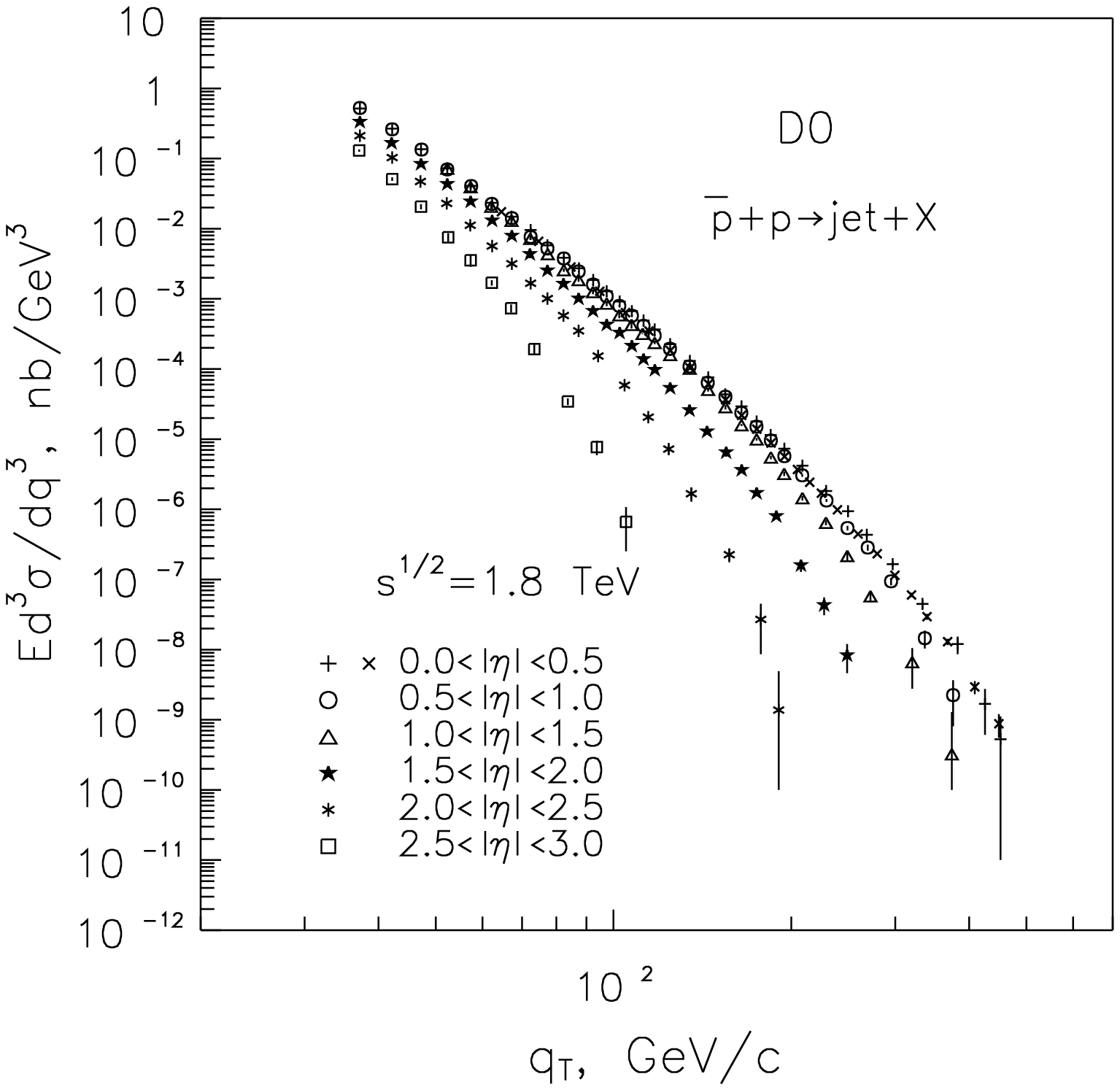}{}}
\hspace*{3cm}
\parbox{5cm}{\epsfxsize=5.cm\epsfysize=5.cm\epsfbox[95 95 400 400]
{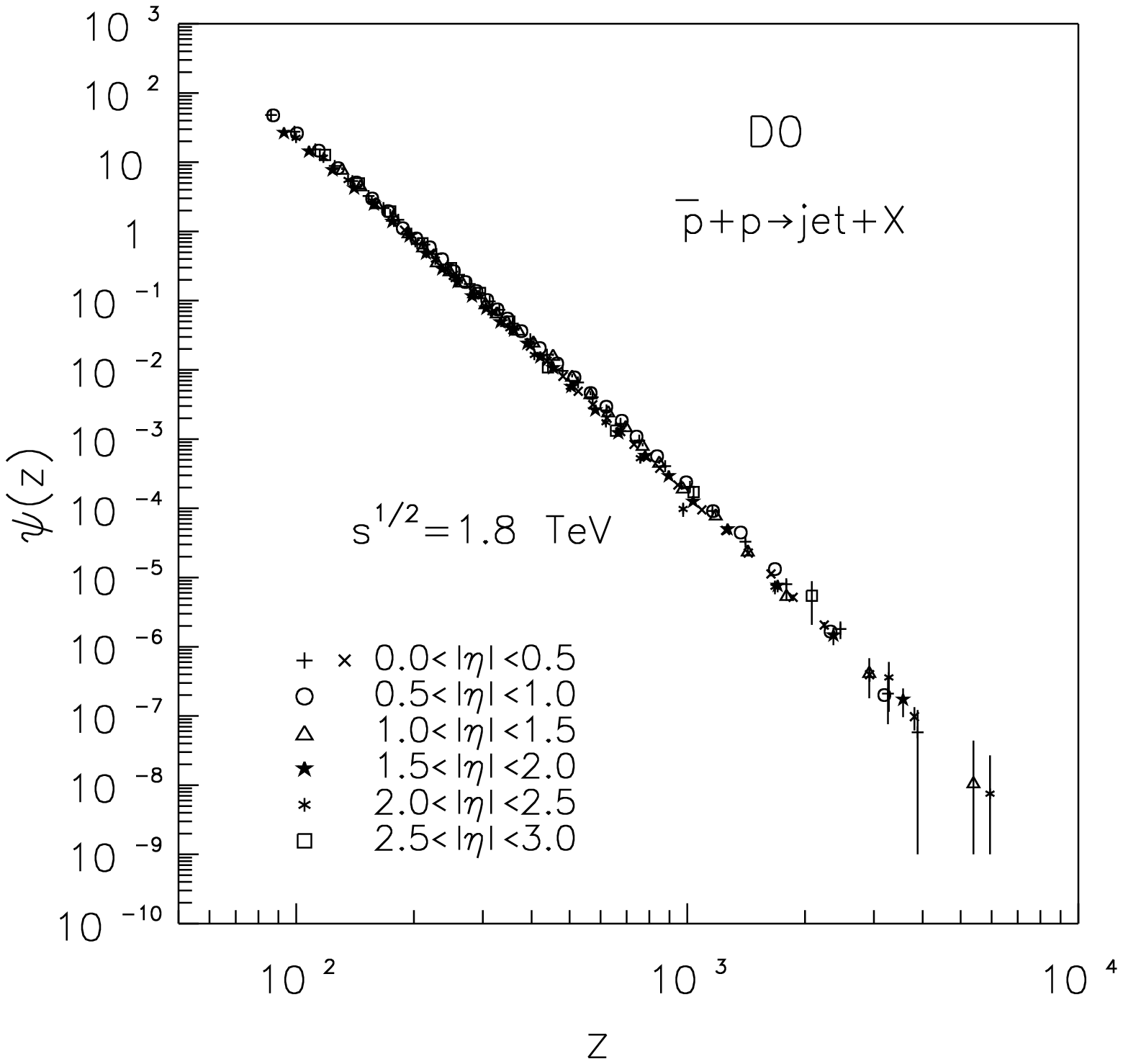}{}}
\vskip -0.5cm
\hspace*{0.cm} a) \hspace*{8.cm} b)\\[0.5cm]
\end{center}

{\bf Figure 11.} (a) The dependence of the inclusive cross section of jet production in $\bar p-p$ collisions
at $\sqrt s  = 1800~GeV/c$ and different rapidity intervals on transverse momentum $q_T$. Experimental data of
the cross section obtained by D0 Collaboration are taken from  \cite{D0A}. (b) The corresponding scaling
function $\psi(z)$.

\vskip 5cm

\begin{center}
\hspace*{-2.5cm}
\parbox{5cm}{\epsfxsize=5.cm\epsfysize=5.cm\epsfbox[95 95 400 400]
{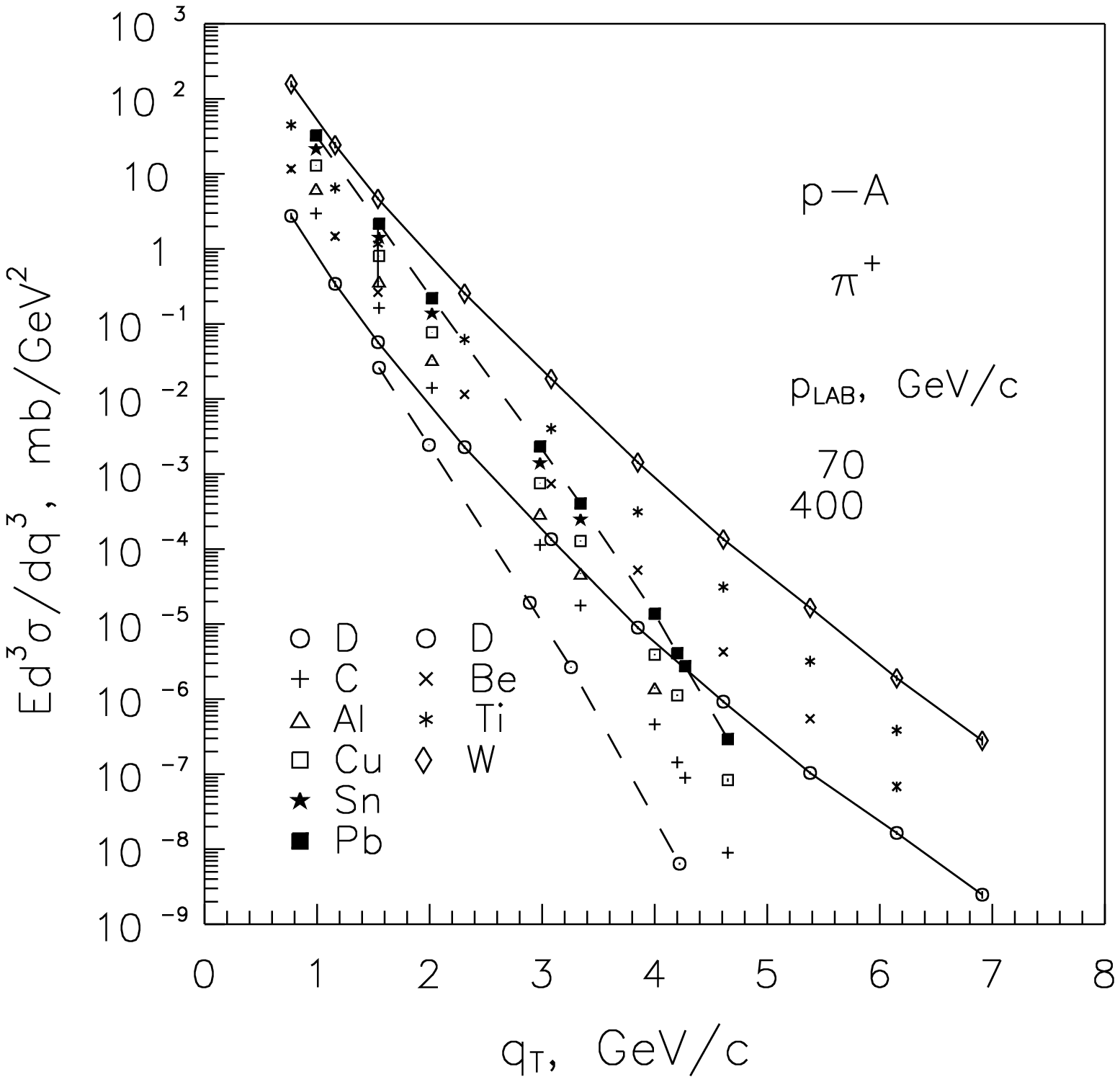}{}}
\hspace*{3cm}
\parbox{5cm}{\epsfxsize=5.cm\epsfysize=5.cm\epsfbox[95 95 400 400]
{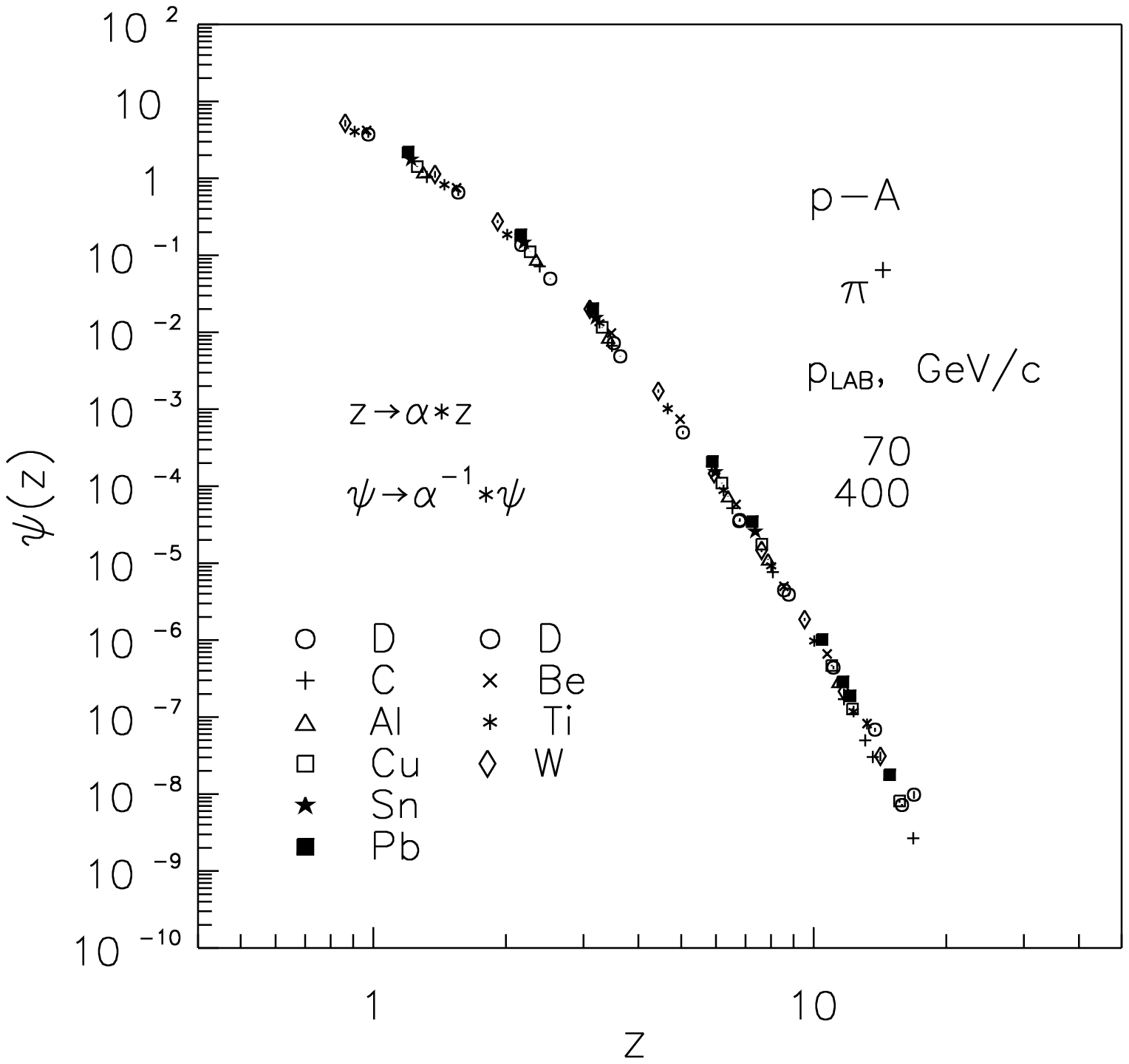}{}}
\vskip -1.cm
\hspace*{0.cm} a) \hspace*{8.cm} b)\\[0.5cm]
\end{center}

{\bf Figure 12.} (a) Inclusive differential cross section
 for the $\pi^+$-mesons produced
in $p-A$ interactions at $p_{lab} = 70, 400~GeV/c$ and in the central region, $\theta_{cm}^{NN} \simeq 90^{0}$
as a function of the transverse momentum  $q_T$.  Solid and dashed lines are obtained by fitting  of the data
for $D,W$ and $D,Pb$, respectively. Experimental data are taken from  \cite{Cronin,Protvino}. (b) The
corresponding  scaling function $\psi(z)$.

\newpage
\begin{minipage}{4cm}

\end{minipage}

\vskip 4cm
\begin{center}
\hspace*{-2.5cm}
\parbox{5cm}{\epsfxsize=5.cm\epsfysize=5.cm\epsfbox[95 95 400 400]
{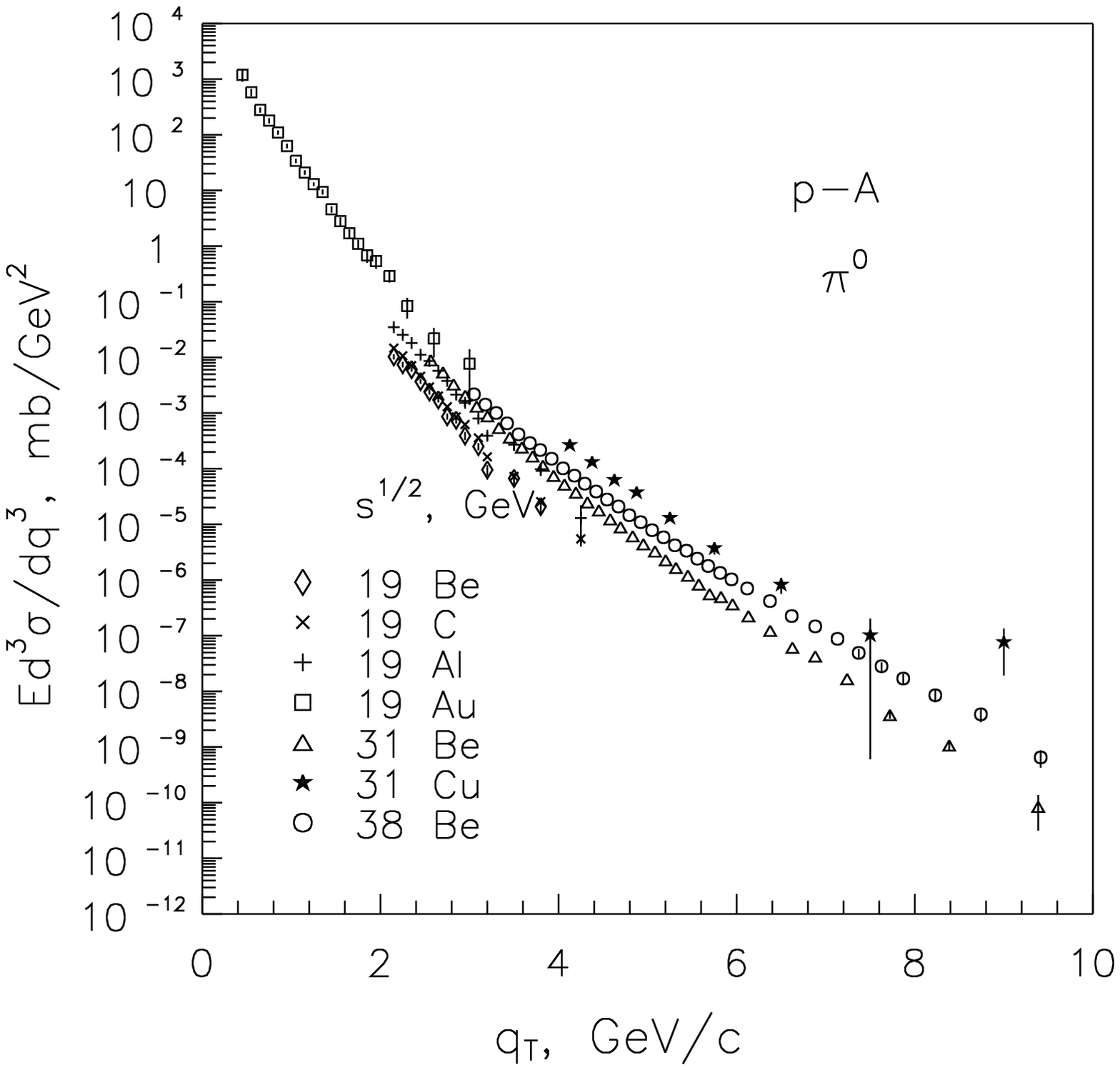}{}}
\hspace*{3cm}
\parbox{5cm}{\epsfxsize=5.cm\epsfysize=5.cm\epsfbox[95 95 400 400]
{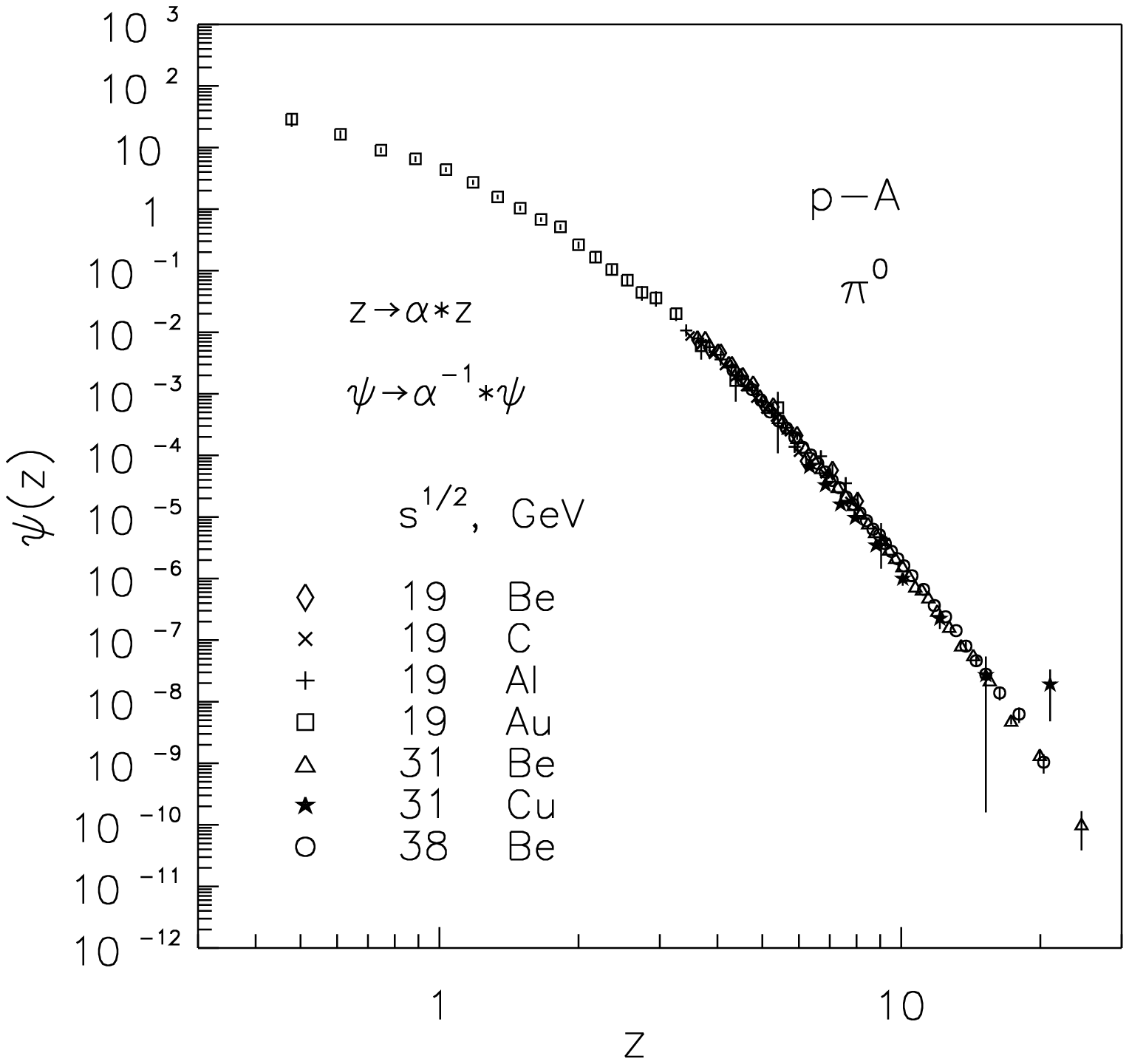}{}}
\vskip -0.5cm
\hspace*{0.cm} a) \hspace*{8.cm} b)\\[0.5cm]
\end{center}

{\bf Figure 13.} (a) Dependence of  the inclusive cross section
 of $\pi^0$-meson  production on transverse
momentum $q_{T}$ in  $p-A$ collisions at $\sqrt s = 19-38~GeV$.
Experimental data are taken from
\cite{Povlis}-\cite{E706}.
(b) The corresponding scaling function $\psi(z)$.

\vskip 5cm

\begin{center}
\hspace*{-2.5cm}
\parbox{5cm}{\epsfxsize=5.cm\epsfysize=5.cm\epsfbox[95 95 400 400]
{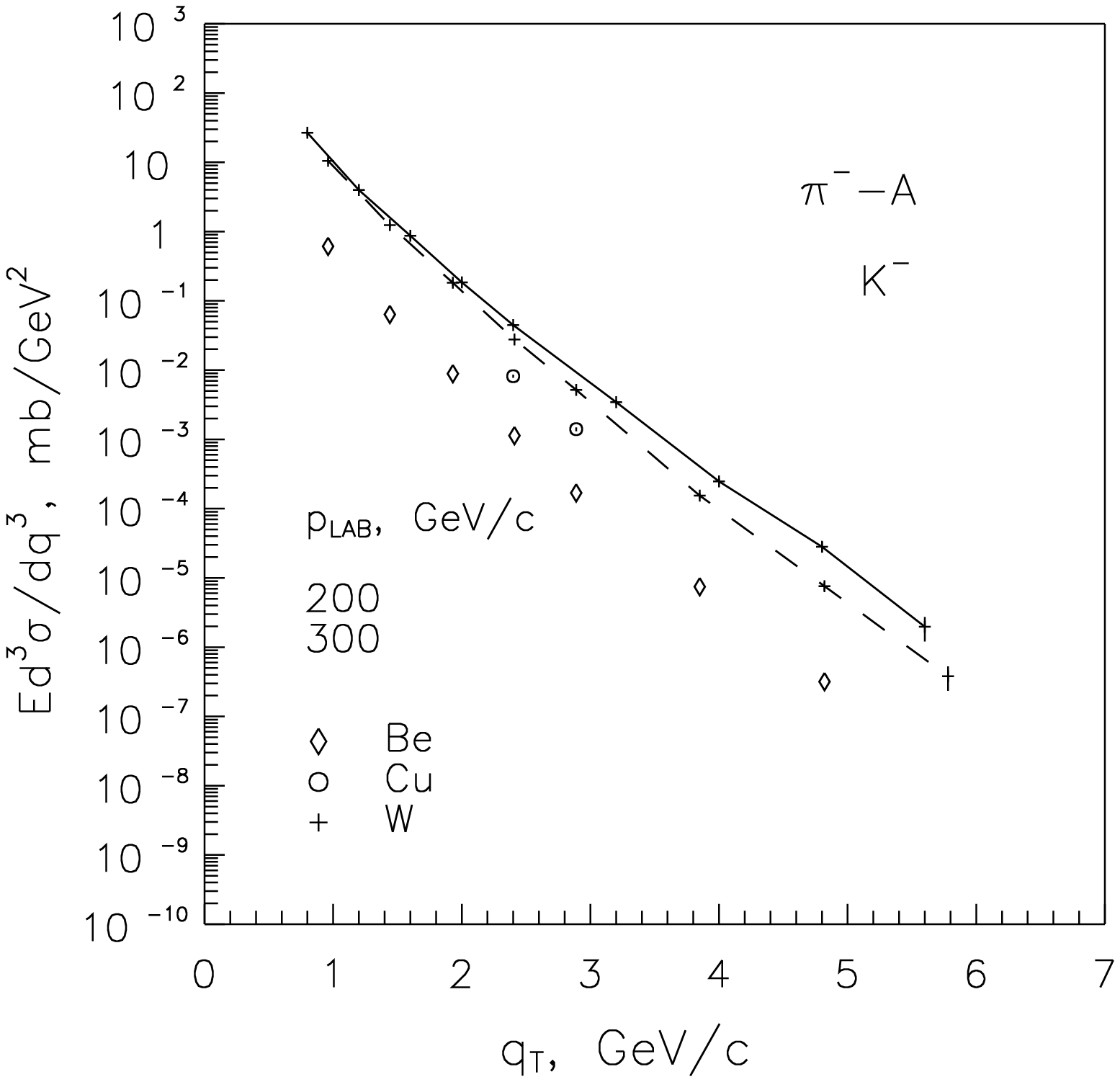}{}}
\hspace*{3cm}
\parbox{5cm}{\epsfxsize=5.cm\epsfysize=5.cm\epsfbox[95 95 400 400]
{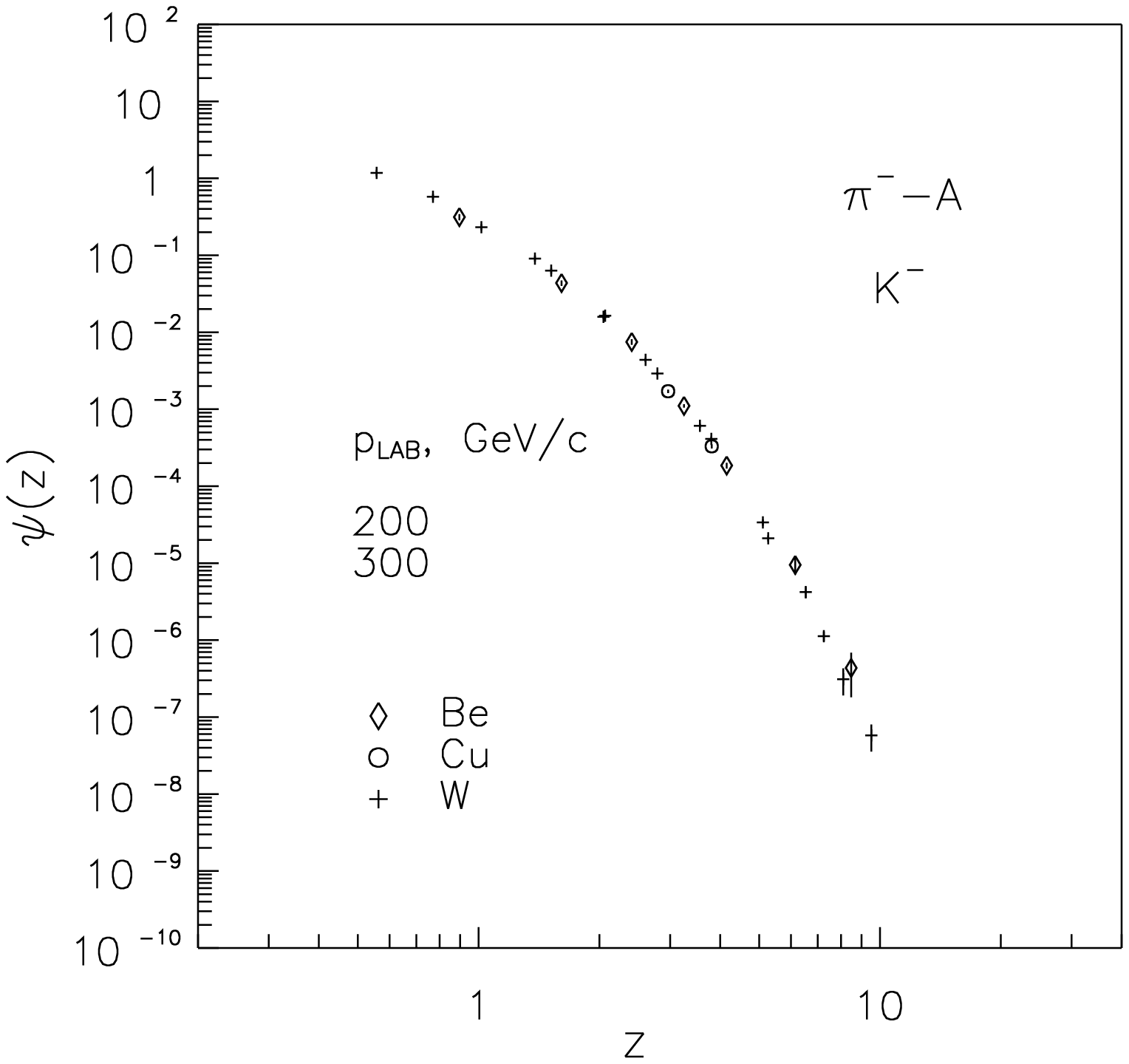}{}}
\vskip -1.cm
\hspace*{0.cm} a) \hspace*{8.cm} b)\\[0.5cm]
\end{center}

{\bf Figure 14.} (a) Dependence
 of  the inclusive cross section of $K^-$-meson  production on transverse
momentum $q_{T}$ in  $\pi^--A$ collisions at $p_{lab} = 200, 300~GeV/c$. Experimental data are taken from
\cite{fris83}. (b) The corresponding scaling function $\psi(z)$.

\newpage
\begin{minipage}{4cm}

\end{minipage}

\vskip 4cm
\begin{center}
\hspace*{-2.5cm}
\parbox{5cm}{\epsfxsize=5.cm\epsfysize=5.cm\epsfbox[95 95 400 400]
{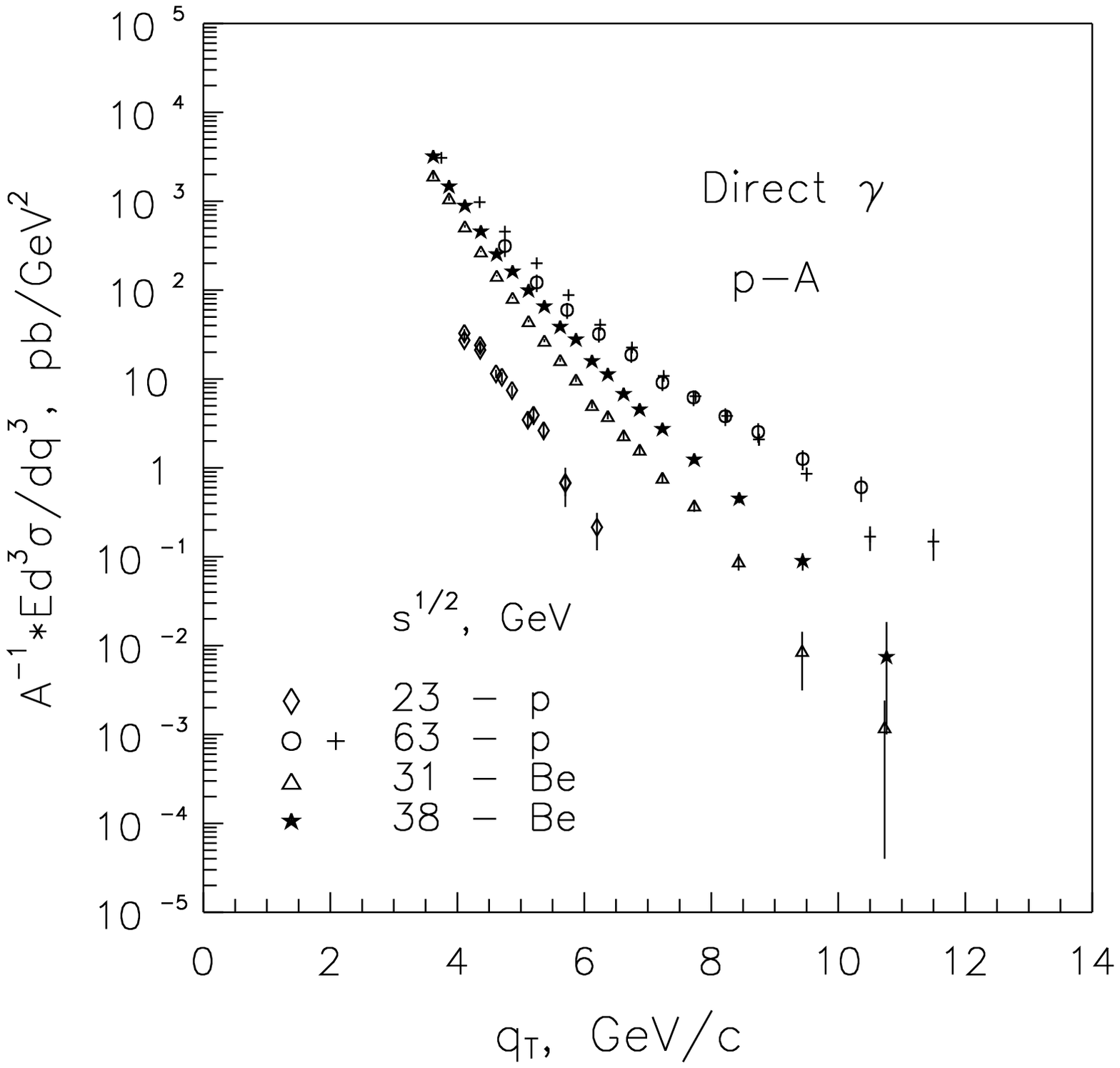}{}}
\hspace*{3cm}
\parbox{5cm}{\epsfxsize=5.cm\epsfysize=5.cm\epsfbox[95 95 400 400]
{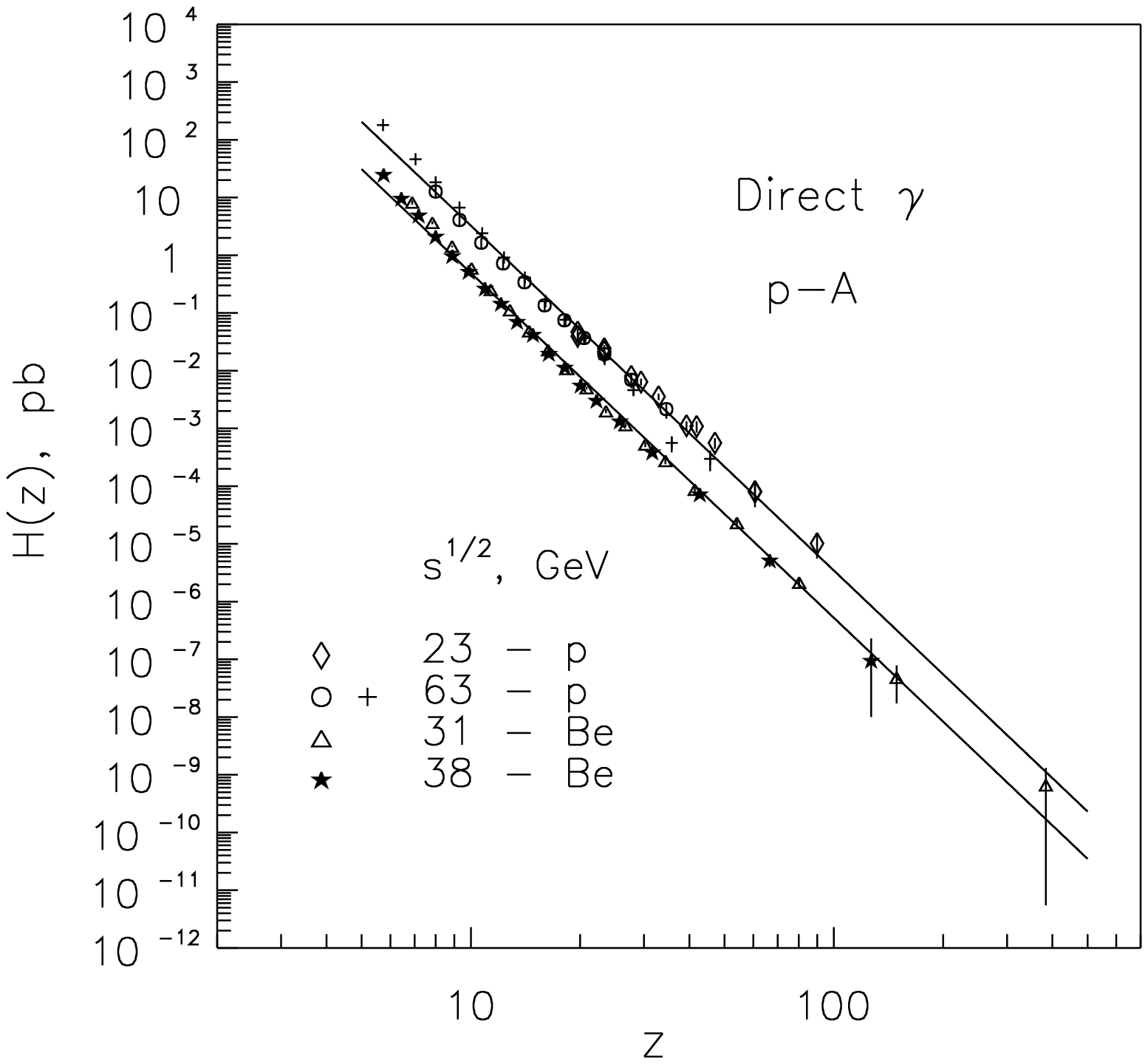}{}}
\vskip -0.5cm
\hspace*{0.cm} a) \hspace*{8.cm} b)\\[0.5cm]
\end{center}

{\bf Figure 15.} (a) Dependence of  the inclusive cross section of direct photon production on transverse
momentum $q_T$ in  $pp$  and $pBe$ collisions. The experimental data on cross section:\ $\diamond$ - WA70
\cite{WA70}, $+$ - R806 \cite{R806}, $\circ$ - R807 \cite{R807}, $\triangle, \star $ - E706 \cite{E706}. (b)
The corresponding scaling function $H(z)$. Solid lines are obtained by fitting the function taken in the form
$H(z)=a_1/z^{a_2}$.

\vskip 5cm

\begin{center}
\hspace*{-2.5cm}
\parbox{5cm}{\epsfxsize=5.cm\epsfysize=5.cm\epsfbox[95 95 400 400]
{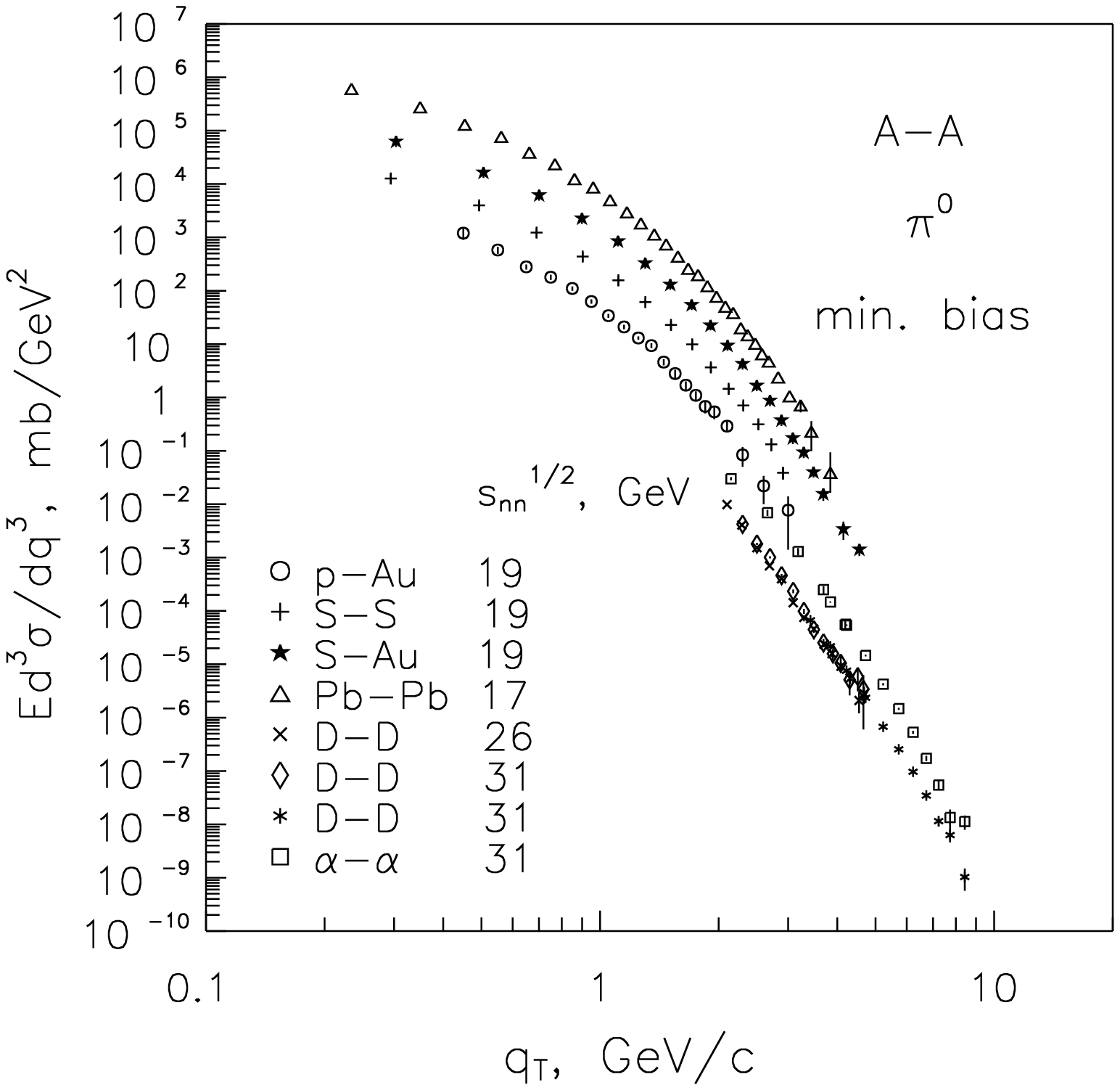}{}}
\hspace*{3cm}
\parbox{5cm}{\epsfxsize=5.cm\epsfysize=5.cm\epsfbox[95 95 400 400]
{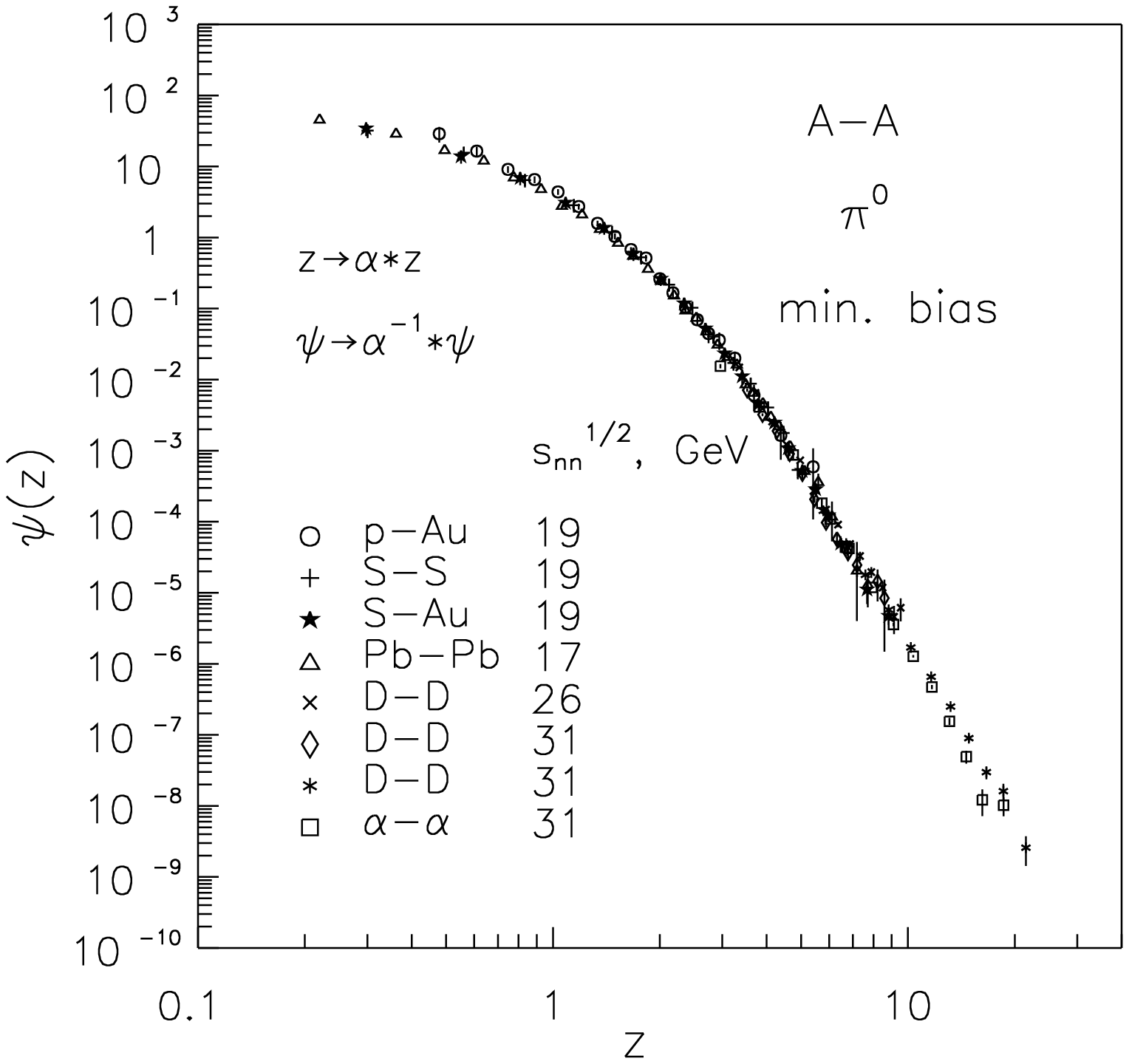}{}}
\vskip -1.cm
\hspace*{0.cm} a) \hspace*{8.cm} b)\\[0.5cm]
\end{center}

{\bf Figure 16.} (a) Dependence of the inclusive cross section of $\pi^0$-meson production on transverse
momentum $q_T$ in minimum bias nucleus-nucleus collisions at $\sqrt s = 17-31~GeV$. Experimental data are
taken from
\cite{Albrecht},\cite{WA80}-\cite{Karabar}. (b) The corresponding scaling function $\psi(z)$.

\newpage
\begin{minipage}{4cm}

\end{minipage}

\vskip 4cm
\begin{center}
\hspace*{-2.5cm}
\parbox{5cm}{\epsfxsize=5.cm\epsfysize=5.cm\epsfbox[95 95 400 400]
{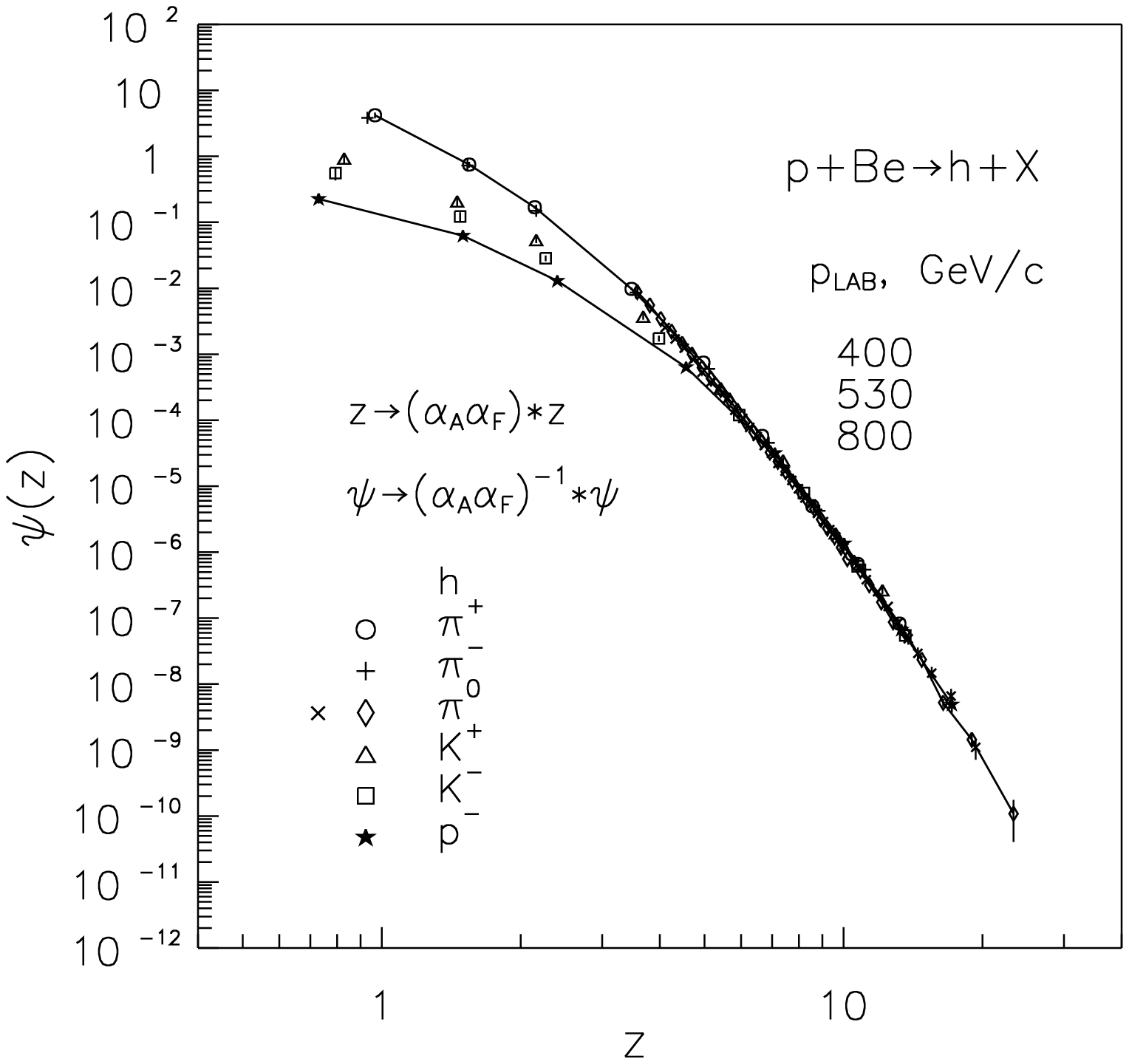}{}}
\hspace*{3cm}
\parbox{5cm}{\epsfxsize=5.cm\epsfysize=5.cm\epsfbox[95 95 400 400]
{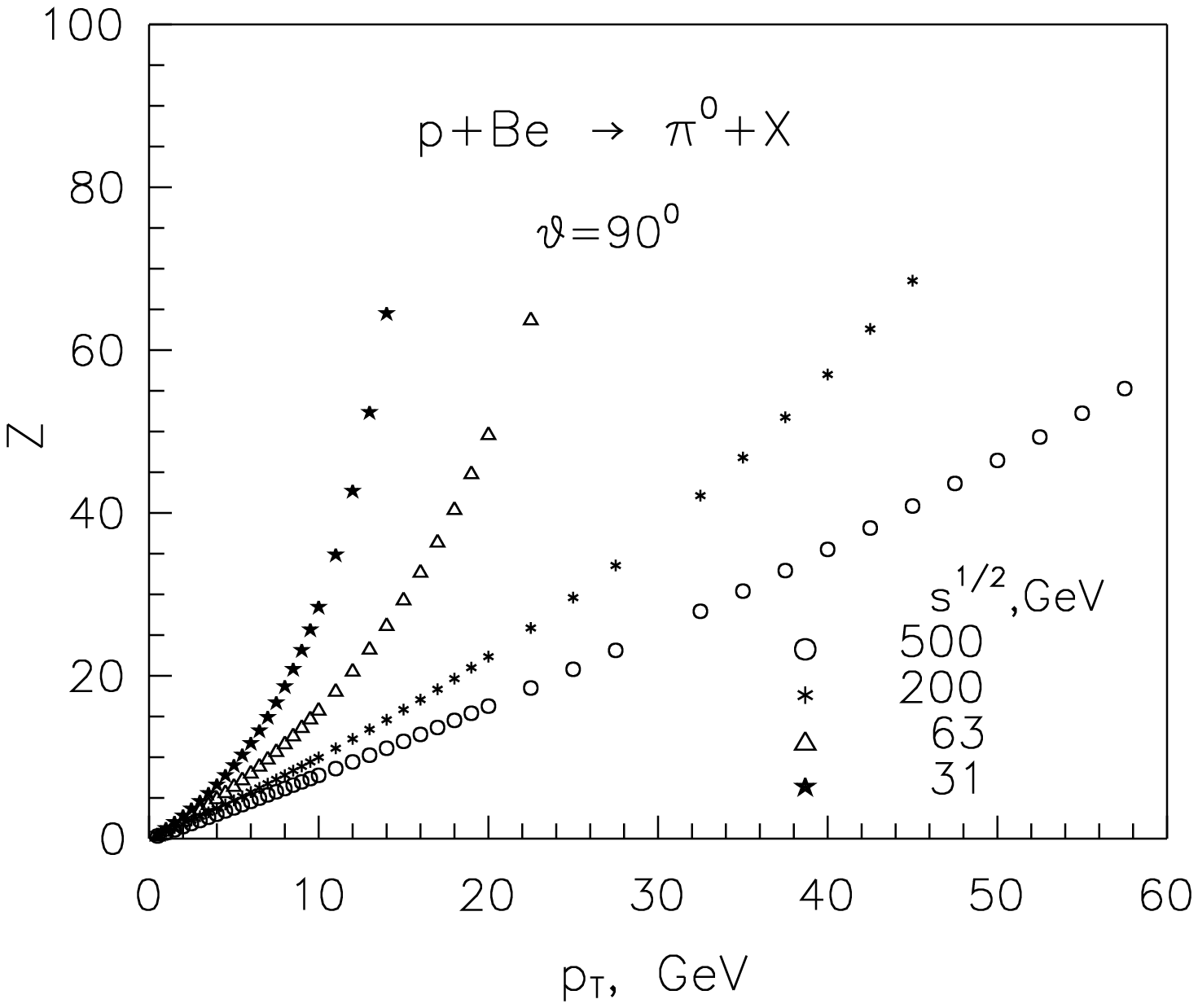}{}}
\vskip -0.5cm
\hspace*{0.cm} a) \hspace*{8.cm} b)\\[0.5cm]
\end{center}

{\bf Figure 17.}  (a) Scaling function $\psi(z)$ for $\pi^{\pm,0}, K^{\pm}, \bar p$ particles produced in
$p-Be$ collisions at $p_{LAB}=400, 530 $ and $800~GeV/c$. Experimental data are taken from \cite{Cronin,E706}.
(b) Dependence of the variable $z$ of $\pi^0$-mesons produced in $p-Be$ collisions on transverse momentum
$q_{T}$ at different energy $\sqrt s$ and $\theta_{cm}^{NN} \simeq 90^0$. Points, $\star - 31~GeV$ , $\diamond
- 63~GeV$, $\circ -200~GeV$,  $+ - 500~GeV$, are the calculated results.

\vskip 5cm

\begin{center}
\hspace*{-2.5cm}
\parbox{5cm}{\epsfxsize=5.cm\epsfysize=5.cm\epsfbox[95 95 400 400]
{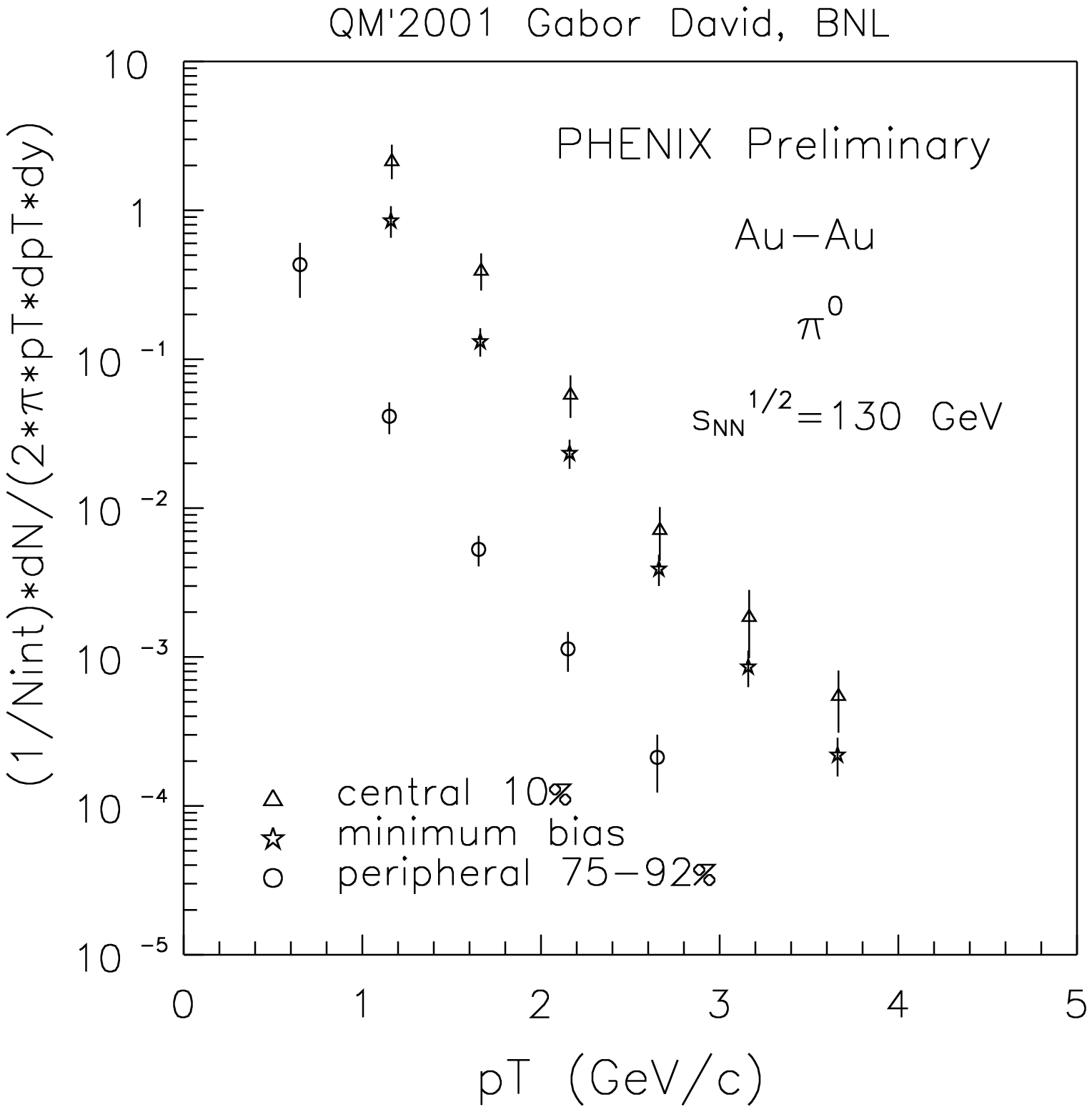}{}}
\hspace*{3cm}
\parbox{5cm}{\epsfxsize=5.cm\epsfysize=5.cm\epsfbox[95 95 400 400]
{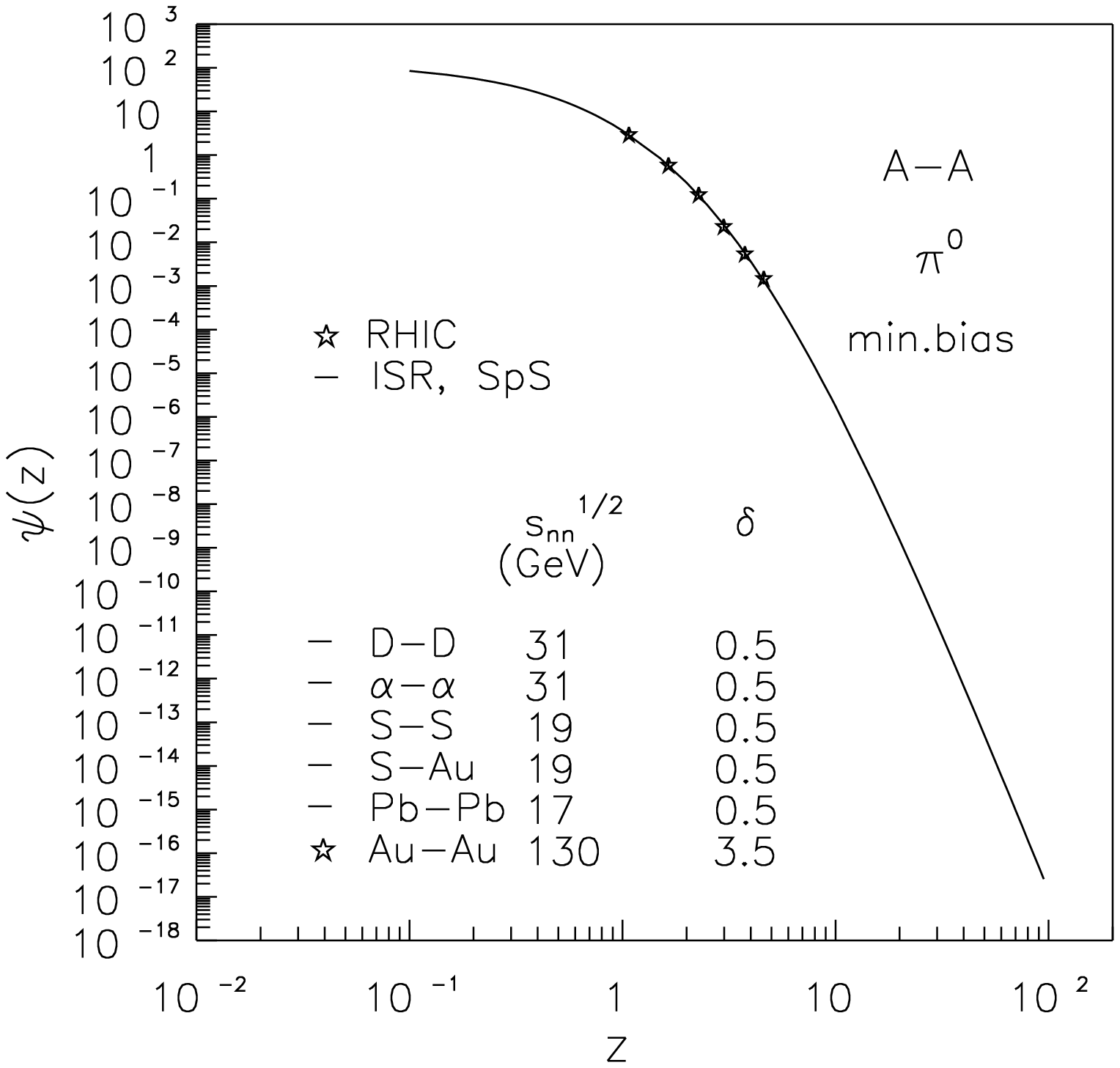}{}}
\vskip -1.cm
\hspace*{0.cm} a) \hspace*{8.cm} b)\\[0.5cm]
\end{center}

{\bf Figure 18.} (a) Spectra of  $\pi^0$-meson production in $Au-Au$ collisions at $\sqrt s_{NN} = 130~GeV$.
Experimental data are obtained by PHENIX Collaboration at RHIC
 \cite{Gabor}.
(b) Scaling function $\psi(z)$ of $\pi^0$-mesons produced in nucleus-nucleus
    collisions  at ISR, SpS and RHIC ($\star $).
The solid line ($ - $) are obtained by fitting of experimental data taken from \cite{Albrecht},
\cite{WA80}-\cite{Karabar}.

\end{document}